\documentclass[aps,prd,nofootinbib,superscriptaddress,floatfix,twocolumn]{revtex4-2}

\usepackage{amsmath,amsfonts,amssymb,color,graphicx,bm,mathrsfs}
\usepackage{xcolor}
\usepackage{mathptmx}
\usepackage{physics}
\usepackage{framed}
\usepackage{float}
\usepackage[normalem]{ulem}
\usepackage{orcidlink}
\usepackage{hyperref}
\usepackage{subcaption}

\begin{document}
\title{\textbf{Detector's response to coherent Rindler and Minkowski photons}}
\author{Pradeep Kumar Kumawat\orcidlink{0009-0007-6535-3869}}
\email{pradeep.kumawat@iitg.ac.in}
\affiliation{Department of Physics, Indian Institute of Technology Guwahati, Guwahati 781039, Assam, India.}
\author{Dipankar Barman\orcidlink{0000-0001-9040-2529}}
\email{dipankar1998@iitg.ac.in}
\affiliation{Department of Physics, Indian Institute of Technology Guwahati, Guwahati 781039, Assam, India.}
\affiliation{Wilczek Quantum Center, Shanghai Institute for Advanced Studies,
Shanghai, 201315, China}
\affiliation{University of Science and Technology of China, Hefei, 230026, China}
\author{Bibhas Ranjan Majhi\orcidlink{0000-0001-8621-1324}}
\email{bibhas.majhi@iitg.ac.in}
\affiliation{Department of Physics, Indian Institute of Technology Guwahati, Guwahati 781039, Assam, India.}

\begin{abstract}
We observe that the transition probability in a static two-level quantum detector interacting with a coherent Rindler field mode differs from that of the Rindler detector interacting with a coherent Minkowski field mode.  The situation does not change in the quantum detector's response in the semiclassical limit of the field state.  This we investigate in $(1+1)$ and $(3+1)$-spacetime dimensions. Interestingly, in $(1+1)$-dimensions, the transition probabilities of the ``classical'' detector in the semiclassical limit of the field state for these two scenarios appear to be identical when the field mode and detector frequencies are taken to be the same. However, in $(3+1)$ dimensions, the detector transition probabilities calculated under the large-acceleration condition do not exhibit such a signature. The implications of these observations are also discussed.

\end{abstract}
\maketitle	

\section{Introduction}
A uniformly accelerated observer perceives the inertial  Minkowski vacuum to be thermally populated with a temperature proportional to the acceleration, known as the Unruh effect \cite{Unruh:1976db}. Einstein's equivalence principle (EEP) bridges this to the Hawking effect \cite{HAWKING:1974us,Hawking1975} where a static observer in black hole spacetime finds the freely falling observer's vacuum (known as the Hartle-Hawking vacuum) to be thermal. These semi-classical results are expected to be very important in illuminating the theory of ``quantum gravity." Therefore, a clear understanding of these effects as well as the validity of the equivalence principles at the quantum regime \cite{Singleton:2011vh,Zych:2015fka,Paunkovic:2022flx} remains a thriving and challenging area in recent times. A freely falling observer towards the black hole horizon becomes excited through an interaction with the field mode defined in the static frame, thereby mimicking the consequence of classical EEP for a quantum system.  One of the remarkable observations in this direction is that a static atom becomes excited by interacting with accelerated field modes \cite{PhysRevLett.121.071301, Louko:2007mu}. In $(1+1)$-dimensions, this particular transition probability is identical to that of an accelerated atom under the exchange of field and atom frequencies, manifesting the quantum equivalence in their classical relative motions \cite{PhysRevLett.121.071301,Fulling:2018lez} (for recent developments, see \cite{Das:2023cfu,Barman:2024dql,Kumawat:2024kul}).

Therefore, understanding the Rindler field with respect to the Minkowski frame is equally important, yet largely lacking in the literature. Analysis in the Rindler vacuum was first established in \cite{Louko:2007mu} and later explored in \cite{PhysRevLett.121.071301}. Although the motions of the atom with respect to the field frame in the two scenarios are referred to as classically analogous, there exist various crucial differences.    
(i) The accelerated atom can see the Rindler horizon, which restricts it from accessing the regions outside the right Rindler wedge (RRW) and therefore interacting only within a sub-region of field space. It is like watching the sea (Minkowski) from a pebble (Rindler) within it. On the other hand, the static atom can access the whole Minkowski spacetime (with no horizon), and so the second scenario is simply like watching a pebble from the sea. Hence, from the perspective of an atom, it is expected that the response to the Minkowski field is different from that to the Rindler field. 
(ii) The fixed inertial frequency ($\omega_M$) of the Minkowski mode function ($u_{k}\sim e^{-i\omega_M t+ik x}$) is perceived by the accelerated atom as $\omega_M e^{-a\tau}$ in the atom's proper time ($\tau$) (see section $14.3$ of \cite{book:PadmanabhanGrav}). Whereas, the same of the Rindler mode function ($u_{k}\sim e^{-i\omega_R \eta+ik\xi}$) is perceived by the inertial atom as $\omega_R/a t$ in the atom's proper time ($t$). Since both horizon and frequency modulation of field modes play a crucial role in the atomic transition (see e.g. \cite{book:PadmanabhanGrav} and \cite{Singh:2013pxf}), clearly these two scenarios are different. The Vacuum state of the field is a purely quantum concept. Therefore, the choice of the field state as the vacuum state in the previous studies \cite{PhysRevLett.121.071301,Louko:2007mu} is not suitable for exploring such crucial, distinct classical features. In particular, the analysis in Ref.~\cite{Louko:2007mu} considered a finite-duration interaction regulated by a smooth switching function, in which transient effects arising from the switching process play an important role.

The coherent field state corresponds to the minimum-uncertainty in position and momentum. Further, the expectation values in this state follow the corresponding classical field evolution, and its coherent amplitude is directly related to the amplitude of the classical field. Therefore, it is the quantum state that most closely resembles a classical field configuration \cite{PhysRevD.38.2457} and may provide a more effective framework for understanding these relative perspectives of the frames. We consider a setup involving a two-level point-like atom coupled to a massless, single-mode, real scalar field \cite{PhysRevLett.121.071301, Kumawat:2024kul, Barman:2024dql}. Then the transitions in the systems will have a classical counterpart along with the (quantum) vacuum contribution \cite{PhysRevD.38.2457} (for other analogous situations see  \cite{Lochan:2014xja, Barman:2022xht}). Since the frequency of a Minkowski field mode, as perceived by an accelerated observer, is different from that of the corresponding Rindler field modes perceived by a static observer, for both classical and quantum fields, the semiclassical limit of the transition probabilities in these two scenarios is expected to exhibit distinctive properties. This allows us to quantify how classicality manifests in a Minkowski coherent field state as observed from the Rindler frame and in a Rindler coherent field state as observed from the Minkowski frame.

We find both in $(1+1)$ and $(3+1)$-dimensions that the excitation and de-excitation probabilities can be decomposed into contributions from the field vacuum state and an additional part due to the presence of excited states in the coherent state, namely the classical part. Contrary to the detector response to the vacuum state analysis, the transition probabilities of accelerating and static atoms do not show any symmetry under the exchange of field and atom frequencies. 
Even the excitation to de-excitation ratios do not reflect any trace of this symmetry, which was present for the vacuum case \cite{PhysRevLett.121.071301, Kumawat:2024kul, Barman:2024dql}. Surprisingly, this mismatch arises from the field's classical contributions. Imposing the semiclassical limit on the field state, one finds that the vacuum part of the transitions in either scenario, as expected, leads to a vanishing contribution. However, the additional contributions (the classical ones) of the transition probabilities survive. In these two scenarios, these terms do not show any such symmetry. This observation is true irrespective of the presence or absence of the reflecting boundary (mirror) in the spacetime. 

The analysis is conducted so that the atom's classical trajectories (either uniformly accelerated or static) are unaffected by the classical value of the field. Therefore, the non-vanishing part of the transition probabilities is purely the classical response of the field on the microstructure of the atom. 
Thus, independent observations of such spacetime dimensions have significant implications \cite{PhysRevLett.121.071301, Barman:2024dql, Kumawat:2024kul}. 
(a) The loss of symmetry between the transition probabilities, even at the semiclassical field limit, indicates that, in general, the response of an accelerated quantum system (here, it is an atom) to the Minkowski field is completely different from that of a static one to the Rindler field. This was not illuminated by the analysis based solely on the vacuum state. 
(b) Reflection of symmetry between the classical trajectories within the transition probabilities is dependent on the background field state.
(c) While the Unruh and Hawking effects are a result of inequivalence of vacuum states in different frames, the present analysis implies that {\it the classical response of the field to a quantum system depends on the choice of respective frames of the field and atom.} 

In addition to the field-semiclassical limit, we analysed the transition probabilities by further imposing the semiclassical limit for the atomic configuration. This accounts for the classical atom's response to the classical field. Such analysis is motivated by the following fact. The classical relative motions of the atom in two scenarios are analogous, and so one can expect that the response of the classical atom to the classical field may be a close counterpart of the reflection of such analogy in motion. To investigate the issue, we consider a very simple situation for which the frequencies of the atom and the field mode are equal (the resonance condition). It is observed that the double-classical response functions in $(1+1)$-dimensions exhibit the symmetry for two different motions of the atom. However, the $(3+1)$-dimensional case is more subtle, and in the region of applicability of the results, we do not find any such symmetry.  

This analysis sheds further light on quantum aspects in a non-inertial frame. 
Moreover, as a result of EEP, this provides new insights into how a freely falling atom towards the black hole horizon will respond to a ``classical'' state (such as a coherent state) of a static observer.

This paper is organized as follows. In Sec. \ref{s2}, we discuss the framework for evaluating the transition
probabilities from the background coherent field state. In Sec. \ref{s3}, we calculate the transition probability of an accelerated atom from the Minkowski field in the presence of a mirror, and that for a static atom from the Rindler field with a mirror in $(1+1)$-dimensions. In Sec. \ref{s4}, we calculate these transition probabilities in  $(1+3)$-dimensions. Then in Sec. \ref{s5}, we take the semiclassical limits on the field and detector systems to analyze the equivalence between the two scenarios. Finally, we conclude with our findings in Sec. \ref{s6}. At the end, we include five appendices to present details of the calculations.

\section{Set-up for coherent field state}\label{s2}

In this section, we review the coherent-state description for our single-mode, massless, real quantum field, and then set up the formalism to calculate the transition probability of a two-level atom locally interacting with the background scalar field when the field is in a coherent state.

The massless scalar field for a specific momentum $k$ (corresponds to ``scalar field'' as called in \cite{PhysRevLett.121.071301}) is known as
\begin{eqnarray}
    \hat{\phi}_k(x)= \sqrt{\hbar_f}\Big(u_{k}(x)\hat{a}_{k}+u_{k}^{\star}(x)\hat{a}_{k}^{\dagger}\Big)\,.
\end{eqnarray}
Here $\hat{a}_{k}$ and $\hat{a}_{k}^{\dagger}$ are annihilation and creation operators, and $u_{k}$ is the mode function. The vacuum is defined as $\hat{a}_{k}|0_{u}\rangle=0$. Here $\hbar_f$ is the Planck constant. The suffix $f$ in the Planck constant stands for the `field'. To specify the quantum structure of the detector atom, we will use the subscript $d$. This is crucial as we are interested in investigating the semiclassical response of the field on the quantum states of the atom. In particular, we denote $\hbar_f\to 0$ as the semiclassical limit of the quantum field while the other Planck constant ($\hbar_d$) remains constant. A general coherent state of the field with coherent amplitude $\alpha_{k}$ is
\begin{eqnarray}
    |\psi_{c}\rangle= \hat{D}(\alpha_{k})|0_{u}\rangle  \,,
\end{eqnarray}
where $\hat{D}(\alpha_{k})=\exp \{ \alpha_{k}(\hat{a}_{k}^{\dagger}-\hat{a}_{k})\}$ is the displacement  operator. The coherent states $|\psi_{c}\rangle$ are eigenstates of the annihilation operator, i.e., $\hat{a}_k|\psi_{c}\rangle=\alpha_k|\psi_{c}\rangle$.

In this study for simplicity of the analytical analysis, we consider a quantum field restricted to a single-frequency mode, similar to the field model considered in Ref.~\cite{PhysRevLett.121.071301}. Physically, such a mode may be regarded as an effective wave-packet mode constructed from a narrow frequency distribution, sharply peaked around a chosen frequency. Single-frequency mode-based field models and their relation to the single-mode approximation were extensively discussed in Ref.~\cite{Bruschi:2010mc}. In particular, that work examined the validity of the single-mode approximation commonly employed in earlier studies of entanglement in non-inertial frames. For example, a maximally entangled state constructed using two single-frequency Minkowski modes corresponding to two Minkowski observers. Later, one observer undergoes uniform acceleration. In this case, as has been shown in \cite{Bruschi:2010mc}, a single-frequency Minkowski mode cannot, in general, be identified with a single-frequency Unruh mode, as assumed in the so-called single-mode approximation. However, a single-frequency Minkowski (Rindler) mode generally decomposes into a superposition of positive- and negative-frequency modes with respect to a Rindler (Minkowski) frame. In our analysis, this frequency mixing precisely leads to a nonvanishing detector response. In particular, a uniformly accelerated detector responds to the Minkowski vacuum (well-known as the Unruh effect), whereas a Minkowski static detector also responds to the Rindler vacuum. In this sense, a direct use of the relation between single-frequency Minkowski and Unruh modes has not been done here.

A two-level atom interacts with the quantum field through the interaction Hamiltonian, modelled as $\lambda\hat{\phi}_k\hat{m}$. Here, the monopole operator of the atom in its proper time $\tau$ is: $\hat{m} = \sqrt{\hbar_d}(\hat{A}e^{-i\Omega \tau} + \hat{A}^\dagger e^{i\Omega\tau})$ (here $\hat{A}$ and $\hat{A}^\dagger$ are the ladder operators of the atom)\cite{PhysRevLett.121.071301}. This type of model has also been considered in \cite{Scully:2017utk,Chakraborty:2019ltu}. Here $\Omega$ denotes the atomic energy gap. In the limit $\hbar_d\to 0$, the energy levels become continuous and the atom behaves as a ``classical'' atom. Note that the introduction of $\hbar_d$ and $\hbar_f$ in the interaction Hamiltonian does not imply distinct fundamental constants, but rather serves as a convenient notation to distinguish between detector and field contributions and to take their (semi)classical limits independently. The atom's transition probability from ground to excited states is (up to $1^{\text{st}}$-order perturbation theory in amplitude) \cite{PhysRevD.38.2457}
\begin{eqnarray}
 P_{\omega}(\Omega)= P_{vac}(\Omega)+ P_{\alpha}(\Omega);
\end{eqnarray}
where, 
\begin{eqnarray}
\label{p-vac}
    &&P_{vac}(\Omega)=\lambda^{2}\hbar_f\hbar_d\Big|\frac{-i}{\hbar_d}\int_{-\infty}^{\infty}d\tau\;  e^{-i\Omega\tau}   u_{k}(x)\Big|^{2}\,;\\\label{p-c}
    &&P_{\alpha}(\Omega)= \lambda^{2}\alpha_{k}^{2}\hbar_f\hbar_{d}\Big|\frac{-i}{\hbar_d}\int_{-\infty}^{\infty}d\tau\; e^{-i\Omega\tau}   (u_{k}(x)+u_{k}^{\star}(x))\Big|^{2}~.
       \end{eqnarray}
Similarly, the de-excitation probability from excited state to ground state of the atom is found to be the same expressions with the replacement $\Omega\to -\Omega$.
Since \(u_k(x)+u_k^\star(x)=2\,\mathrm{Re}[u_k(x)]\) is a real function along the detector trajectory, from Eq. (\ref{p-c}), we can observe that under the replacement \(\Omega\to -\Omega\) the integrand becomes the complex conjugate of the original one. Therefore, the quantity inside the modulus changes only by complex conjugation, leaving the modulus squared invariant. Hence, $
P_\alpha(-\Omega)=P_\alpha(\Omega),$ shows that \(P_\alpha\) is an even function of \(\Omega\). The response of the quantum atom to the classical field is obtained by taking $\hbar_f\to 0$ while $\alpha_k^2\hbar_f$ is finite \cite{PhysRevD.38.2457}. Further imposition of the limit $\hbar_d\to 0$ yields the response of ``classical atom'' to the classical field.

\section{Atomic transitions in $(1+1)$-dimensional spacetime}\label{s3}
\begin{figure}[t]

    \centering
    \begin{subfigure}[t]{0.48\linewidth}
    \centering
    \includegraphics[width=1.0\linewidth]{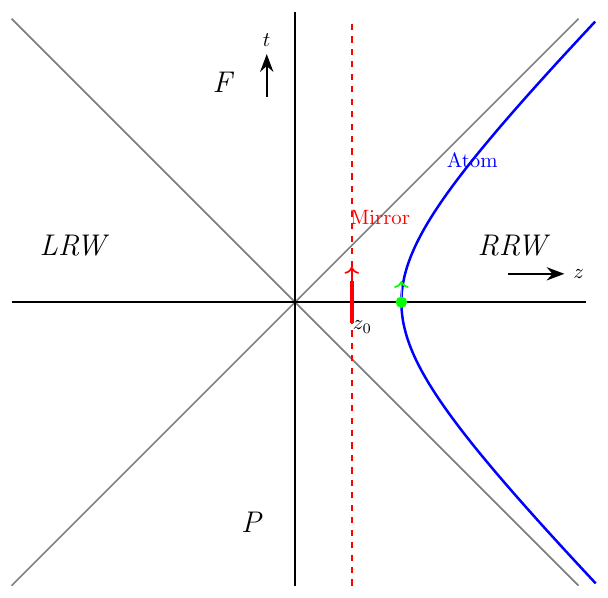}
    \subcaption{}
    \label{fig:Acc_Atom}
    \end{subfigure}
    \hfill
    \begin{subfigure}[t]{0.48\linewidth}
    \centering
    \includegraphics[width=1.0\linewidth]{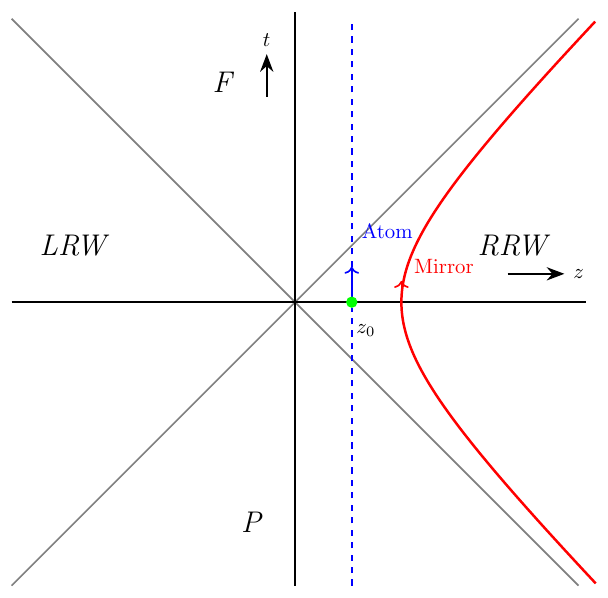}
    \label{fig:Static_Atom}
    \subcaption{}
    \end{subfigure}

    \caption{Spacetime diagrams of the two configurations considered in this work: (a) A uniformly accelerated atom (blue trajectory) in the presence of a static mirror at $z=z_0$ (red dashed line); (b) A static atom at $z=z_0$ (blue dashed line) in the presence of a uniformly accelerated mirror (red trajectory). Here, RRW, LRW, F, and P denote the right Rindler wedge, left Rindler wedge, future wedge, and past wedge, respectively. }
  \label{fig:Atom_mirror}
\end{figure}

We consider a reflecting boundary in $(1+1)$-dimensional spacetime, where the field modes vanish. Here, we consider two scenarios: $(i)$ the atom is accelerating with uniform acceleration, while the mirror is static (see Fig.~(\ref{fig:Atom_mirror} a)), and $ (ii)$ the mirror is uniformly accelerating, while the atom is static in the Minkowski spacetime (see Fig.~(\ref{fig:Atom_mirror} b)).

\subsection{The atom is uniformly accelerating}
The atom is accelerating with uniform acceleration $a$ along the positive Minkowski $z$-axis. The coordinate transformation of this detector can be written in terms of Rindler coordinates ($\eta, \xi$) as
\begin{eqnarray}\label{xtD}
    t=\frac{e^{a \xi}}{a}\sinh{a\eta}~;~~~~
    z=\frac{e^{a\xi}}{a}\cosh{a\eta}\,.
\end{eqnarray}
The Rindler coordinate time ($\eta$) coincides with the detector's proper time ($\tau$) for the $\xi=0$ trajectory. The mirror is kept fixed at $z=z_{0}$. 
In the presence of this mirror, the Minkowski scalar field mode solution is given as $u_{k}^{M}(t,z)=e^{- i\,\omega\, t}(e^{i\,k\,(z-z_{0})}-e^{-i\,k\,(z-z_{0})})/\sqrt{4\pi\omega}$. This Minkowski mode can be expressed along the worldline of the accelerated atom using the coordinate transformation (with $\xi=0$) in Eq. (\ref{xtD}) as 
\begin{equation}\label{mm_withM_1p1D}
    u_{k}^{M}(x(\tau))=\frac{1}{\sqrt{4\pi\omega}}(e^{\frac{ i\omega}{a}e^{-a\tau} }e^{-i\,\omega\,z_{0}}-e^{-\frac{ i\omega}{a}e^{a\tau} }e^{i\,\omega\,z_{0}})\,.
\end{equation}
The detector excitation and de-excitation probabilities in the presence of a mirror from the field vacuum can be evaluated using Eq. (\ref{p-vac}) and (\ref{mm_withM_1p1D}), and are known as \cite{Kumawat:2024kul}
\begin{eqnarray}\label{P-e-wm1}
   && P_{vac}(\Omega)
  =\frac{2\lambda^{2}\hbar_f}{\hbar_{d}\omega a \Omega}\frac{\sin^{2}{(\omega\,z_{0}+\varphi_{1})}}{e^{\frac{2\pi\Omega }{a}}-1} \,;\\
  \label{P-d-wm1}
   && P_{vac}(-\Omega)
  =\frac{2\lambda^{2}\hbar_f}{\hbar_{d}\omega a \Omega}\frac{\sin^{2}{(\omega\,z_{0}-{\varphi}_{1})}}{1-e^{-\frac{2\pi\Omega }{a}}} \,~,
\end{eqnarray}
where $\varphi_{1}=\mathrm{Arg}[(\omega/a)^{i\,\Omega/a}\,\Gamma(-i\,\Omega/a)]$.
 Then, from Eq. (\ref{p-c}) and (\ref{mm_withM_1p1D}), the non-vacuum contributions in the detector excitation probability is obtained as 
\begin{eqnarray}
    P_{\alpha}(\Omega)&=& \frac{\lambda^{2}\,\alpha_{k}^{2}\,\hbar_{f}}{4\pi\omega\,\hbar_{d}} \bigg|\int_{-\infty}^{\infty}d\tau\; e^{-i\Omega\tau}    \big(
      e^{\frac{ i\omega}{a} e^{-a\tau}} e^{-i\omega z_{0}}
      \nonumber
      \\
      &&+ e^{-\frac{ i\omega}{a} e^{-a\tau}}e^{i\omega z_{0}}\,
      - e^{-\frac{ i\omega}{a} e^{a\tau}}e^{i\omega z_{0}}- e^{\frac{ i\omega}{a} e^{a\tau}}e^{-i\omega z_{0}}   \bigl)\bigg|^{2}\,\nonumber\\ 
     &=& \frac{\lambda^{2}\,\alpha_{k}^{2}\,\hbar_{f}}{4\pi\omega a^{2}\,\hbar_{d}} \bigg|\int_{0}^{\infty}du\; u^{\frac{i\Omega}{a}-1}    \big(
      e^{\frac{ i\omega}{a} u} e^{-i\omega z_{0}}+ e^{-\frac{ i\omega}{a} u}e^{i\omega z_{0}}\big)\,
      \nonumber\\
      &&-\int_{0}^{\infty}d\tilde{u}\; \tilde{u}^{-\frac{i\Omega}{a}-1}    \big( e^{-\frac{ i\omega}{a} \tilde{u}}e^{i\omega z_{0}}+ e^{\frac{ i\omega}{a} \tilde{u}}e^{-i\omega z_{0}}   \bigl)\bigg|^{2}\,.
\end{eqnarray}  
To obtain the above expression, we considered the change in the variables as $e^{-a\tau}=u$ and $e^{a\tau}=\tilde{u}$. Then, we used the following integral relation to obtain the final expression:
\begin{eqnarray}\label{GammaR}
\int_{0}^{\infty}x^{\nu}e^{-a x}dx=\frac{\Gamma(\nu+1)}{a^{\nu+1}}\,.
\end{eqnarray}
And obtained the final expression as
\begin{eqnarray}\label{P-a-e-1a}
    &&P_{\alpha}(\Omega)=\frac{\lambda^{2}\,\alpha_{k}^{2}\hbar_f}{\,\hbar_{d}\Omega\,\omega \,a\,\sinh(\frac{\pi\Omega}{a})}\bigg|-e^{\frac{\pi\Omega}{2a}}\sin(\omega z_{0}-{\varphi}_{1}) \big)\,\nonumber\\ 
    &&~~~~~~~~~~~~~~~~~~~~~~~~~~~~~~~~~~~~~\,+e^{-\frac{\pi\Omega}{2a}}\sin(\omega z_{0}+{\varphi}_{1})    \bigg|^{2}\,.
\end{eqnarray}
 Here again, the same phase $\varphi_1$ appeared as in the expression $P_{vac}(\Omega)$.

\subsection{The atom is static}\label{Sec1plus1static}
In this scenario, we consider that the mirror is accelerating uniformly with acceleration $a$ along the $z$-axis, while the atom is stationary at $z=z_{0}$ ($z_0<1/a^{}$). The moving mirror follows the trajectory given by Eq. (\ref{xtD}) in Minkowski space, and in the Rindler space it is fixed at $\xi=0$.  Then the Rinder field mode solution is given as $u_{k}^{R}(\eta,\xi)=e^{- i\,\omega\,\eta }(e^{i\,k\,\xi}-e^{-i\,k\,\xi})/\sqrt{4\pi\omega}$. Here $k$ is the wave vector, and  $\omega$ is the frequency of the field mode ($\omega=|k|$). We investigate a static detector located at this position interacting with a field mode in the right Rindler wedge. For such a detector, the worldline spans three regions of Minkowski spacetime: the past wedge ($t < -z_0$), the right Rindler wedge ($-z_0 < t < z_0$), and the future wedge ($t > z_0$). Therefore, the detector interacts with the field over its entire proper time, and one must consider a field mode that is well-defined along the full trajectory. The vacuum corresponding to a static observer is the Minkowski vacuum, which is defined over the full Minkowski spacetime. Similarly, the vacuum corresponding to a uniformly accelerated frame is the Rindler vacuum (here, the mirror frame), which in our case, is confined to the right Rindler wedge. However, one can construct field modes by appropriately combining the modes belonging to different Rindler wedges so that their vacuum state coincides with the Minkowski one. In particular, combining the right and left Rindler wedge modes yields the well-known Unruh modes, and the Minkowski vacuum acts as its vacuum state \cite{Unruh:1976db}. Likewise, one may combine the future and past wedge modes such that it coincides with the Minkowski vacuum \cite{Olson:2010jy}.
In our case, the observer's trajectory passes successively through the past, right, and future wedges while remaining coupled to the same background quantum field, defined in right Rindler wedge(see Fig.~(\ref{fig:Atom_mirror} b)). In the past (future) wedge, the detector interacts with the right (left)-moving mode obtained by analytically continuing the right Rindler modes into that region. The  Rindler field mode can be expressed in terms of the atom's coordinates (using Eq. (\ref{xtD})) as
\begin{equation}\label{rm_withM_1p1D}
    u_{k}^{R}(t,z_{0})=\frac{[(a(z_{0}-t))^{\frac{i\omega}{a}}\theta(z_{0}-t)-(a(z_{0}+t))^{-\frac{i\omega}{a}}\theta(z_{0}+t)]}{\sqrt{4\pi\omega}}\, ,
\end{equation}
where $\theta$ is the Heaviside step function. The Rindler mode in Eq.~(\ref{rm_withM_1p1D}) is naturally expressed in terms of null coordinates $u = z - t$, and $v = z + t$, which allows a decomposition into right-moving (incident) and left-moving (reflected) components. These components propagate along distinct null directions and originate from different spacetime regions.
It is well known that such modes admit analytic continuation across Rindler horizons: right-moving modes originating in the past wedge extend into the right wedge, while left-moving modes originating in the right wedge extend into the future wedge (see, e.g., \cite{Higuchi:2017gcd, PhysRevLett.121.071301}). A similar argument is also presented in Ref.~\cite{PhysRevLett.121.071301}. Using this property, the mode function in Eq.~(\ref{rm_withM_1p1D}) provides a consistent description along the entire detector trajectory.
Importantly, the Heaviside functions appearing in Eq.~(\ref{rm_withM_1p1D}) implement causality rather than an arbitrary prescription. For a detector at $z = z_0$, the reflected (left-moving) component contributes only for $t > -z_0$, when signals from the mirror can reach the detector, while the incident (right-moving) component contributes only for $t < z_0$, after which it has already passed the detector. Thus, the field experienced by the detector evolves in a manner consistent with causal propagation.

Then, the detector excitation and de-excitation probabilities from the field vacuum can be evaluated using Eq. (\ref{p-vac}) and (\ref{rm_withM_1p1D}), and are known as \cite{Kumawat:2024kul}
\begin{eqnarray}\label{P-e-wm1s}
   && P_{vac}(\Omega)
  =\frac{2\lambda^{2}\hbar_f}{ a \Omega^{2}\hbar_{d}}\frac{\sin^{2}{(\omega\,z_{0}+\varphi_{2})}}{e^{\frac{2\pi\omega }{a}}-1} \,;\\
    \label{P-d-wm1s}
   && P_{vac}(-\Omega)
  =\frac{2\lambda^{2}\hbar_f}{ a \Omega^{2}\hbar_{d}}\frac{\sin^{2}{(\omega\,z_{0}-{\varphi}_{2})}}{1-e^{-\frac{2\pi\omega }{a}}} \,~,
\end{eqnarray}
where $\varphi_{2}=\mathrm{Arg}[(\Omega/a)^{i\,\omega/a}\,\Gamma(-i\,\omega/a)]$.
Now, using (\ref{p-c}) and (\ref{rm_withM_1p1D}), the non-vacuum contributions in transition probability is obtained as 

\begin{eqnarray}
   P_{\alpha}(\Omega)&=&\frac{\lambda^{2}\alpha_{k}^{2}\,\hbar_{f}}{4\pi\omega\,\hbar_{d}}\bigg|\int_{-\infty}^{z_{0}}dt\; e^{-i\Omega t}    \big[(a(z_{0}-t))^{\frac{i\omega}{a}}+(a(z_{0}-t))^{-\frac{i\omega}{a}}\big]\,
   \nonumber
   \\
   &&- \int_{-z_{0}}^{\infty}dt\; e^{-i\Omega t}    \big[(a(z_{0}+t))^{\frac{-i\omega}{a}}+(a(z_{0}+t))^{\frac{i\omega}{a}}\big]\bigg|^{2}\,\nonumber\\ 
  &=&  \frac{\lambda^{2}\alpha_{k}^{2}\,\hbar_{f}}{4\pi\omega a^{2}\,\hbar_{d}}\bigg|\int_{0}^{\infty}dy\; e^{i\Omega (\frac{y}{a}-z_{0})}    \big[y^{\frac{i\omega}{a}}+y^{-\frac{i\omega}{a}}\big]\,
  \nonumber
  \\
  &&- \int_{0}^{\infty}d\tilde{y}\; e^{-i\Omega (\frac{\tilde{y}}{a}-z_{0})}    \big[\tilde{y}^{\frac{-i\omega}{a}}+\tilde{y}^{\frac{i\omega}{a}}\big]\bigg|^{2}\,. 
\end{eqnarray}

To obtain the above expression, we made changes in variables as $a(z_{0}-t)=y$ and $a(z_{0}+t)=\tilde{y}$. Then, using the integral relation given in Eq.~(\ref{GammaR}), we obtain the final expression 

\begin{eqnarray}\label{P-a-e-1s}
 &&   P_{\alpha}(\Omega)=\frac{\lambda^{2}\,\alpha_{k}^{2}\hbar_f}{a\,\Omega^{2}\,\hbar_{d}\,\sinh(\frac{\pi\omega}{a})}\bigg|-e^{\frac{\pi\omega}{2a}}\sin( \Omega\,z_{0}-\varphi_{2})\nonumber\\ &&
~~~~~~~~~~~~~~~~~~~~~~~~~~~~~~~~~~~~+ e^{-\frac{\pi\omega}{2a}}\sin(\Omega\,z_{0}+\varphi_{2})
      \bigg|^{2}\,.
\end{eqnarray}

For completeness and more visibility on the implications, the corresponding results in free spacetime (i.e. without a boundary or mirror) 
have also been obtained in Appendix \ref{Appendix_NM_1}. We summarize all these below in tabular form (see tables \ref{T1} and \ref{T2}) in order to compare different scenarios.  
\begin{widetext}
\begin{center}
 \begin{table}[h!]
 \caption{The transition probability of the atom with a mirror in the $(1+1)$ dimensions }
     \label{T1} 
 {
    \centering
    \begin{tabular}{ |c|c|c|  }
 \hline

 \multicolumn{3}{|c|}{Transition Probability in $(1+1)$ Dimensions with Mirror $( P_{\omega}(\Omega)= P_{vac}(\Omega)+ P_{\alpha}(\Omega))$} \\
 \hline    Atom accelerated/static & Vacuum contribution  & Non-vacuum contribution \\
 \hline  Accelerated & 
    $\begin{aligned}
    &{P}_{vac}(\Omega) 
    = \frac{2\,\lambda^2\hbar_f}{a\,\Omega\,\omega\,\hbar_{d}\,}\,\frac{\sin^2{(\omega\,z_{0}+\varphi_{1})}}{e^{\frac{2\pi\,\Omega}{a}}-1}\\
    &{P}^{}_{vac}(-\Omega)=\frac{2\,\lambda^2\hbar_f}{a\,\Omega\,\omega\,\hbar_{d}\,}\,\frac{\sin^2{(\omega\,z_{0}-\varphi_{1})}}{1-e^{-\frac{2\pi\,\Omega}{a}}}
    \end{aligned}$ &    
        $\begin{aligned}
         &{P}^{}_{\alpha}(\pm\Omega)=\frac{\lambda^{2}\,\alpha_{k}^{2}\hbar_f}{4\,\Omega\,\omega \,a\,\hbar_{d}\,\sinh(\frac{\pi\Omega}{a})}\bigg|-e^{\frac{\pi\Omega}{2a}}\sin(\omega z_{0}-{\varphi}_{1}) \big)\,\nonumber\\ 
    &~~~~~~~~~~~~~~~~~~~~~~~~~~~~~~~~~~~~~\,+e^{-\frac{\pi\Omega}{2a}}\sin(\omega z_{0}+{\varphi}_{1})    \bigg|^{2}
         \end{aligned}$\\ \cline{1-3}
    Static & $\begin{aligned}
    &P^{}_{vac}(\Omega)=\frac{2\,\lambda^2\hbar_f}{a\,\Omega^2\,\hbar_{d}\,}\,\frac{\sin^2{(\Omega\,z_{0}+\varphi_{2})}}{e^{\frac{2\pi\,\omega}{a}}-1}   \\
    &P_{vac}(-\Omega)=\frac{2\,\lambda^2\hbar_f}{a\,\Omega^2\,\hbar_{d}\,}\,\frac{\sin^2{(\Omega\,z_{0}-{\varphi}_{2})}}{1-e^{-\frac{2\pi\,\omega}{a}}}\end{aligned}$&   
    $\begin{aligned}
        &{P}^{}_{\alpha}(\pm\Omega) =\frac{\lambda^{2}\,\alpha_{k}^{2}\hbar_f}{a\,\Omega^{2}\,\hbar_{d}\,\,\sinh(\frac{\pi\omega}{a})}\big|-e^{\frac{\pi\omega}{2a}}\sin( \Omega\,z_{0}-\varphi_{2})\\ &
~~~~~~~~~~~~~~~~~~~~~~~~~~~~~~~~~~~~~~+ e^{-\frac{\pi\omega}{2a}}\sin(\Omega\,z_{0}+\varphi_{2})
      \big|^{2}
    \end{aligned}$\\\cline{1-3}
   
 \hline
\end{tabular}

     }
\end{table}

\begin{table}[h!] 
\caption{The transition probability of the atom without a mirror in the $(1+1)$ dimensions. In the expressions below, $\phi$ is defined as $\phi=Arg\big[\big(\Omega/a\big)^{i\omega/a}\;\Gamma\big(-i\omega /a\big)\big]$.}
     \label{T2}
 {
    \centering
    \begin{tabular}{|c|c|c|  }
 \hline
\multicolumn{3}{|c|}{Transition Probability in $(1+1)$ Dimensions $( P_{\omega}(\Omega)= P_{vac}(\Omega)+ P_{\alpha}(\Omega))$} \\
 \hline    Atom accelerated/static & Vacuum contribution  & Non-vacuum contribution \\
 \hline  Accelerated & $\begin{aligned}
 &P_{vac}(\Omega)
  =\frac{\lambda^{2}\hbar_f}{2\omega a \Omega\,\hbar_{d}\,}\frac{1}{e^{\frac{2\pi\Omega }{a}}-1}  \\
  \label{P-d-v1}
&P_{vac}(-\Omega)
  =\frac{\lambda^{2}\hbar_f}{2\omega a \Omega\,\hbar_{d}\,}\frac{1}{1-e^{-\frac{2\pi\Omega }{a}}} \end{aligned}$&    
        $\begin{aligned}
         {P}^{}_{\alpha}(\pm\Omega)&=\frac{\lambda^2\alpha_{k}^{2}\hbar_f}{2\omega a \Omega\,\hbar_{d}\,}\coth{\bigg(\frac{\pi\Omega}{2a}\bigg)}
         \end{aligned}$\\ \cline{1-3}
    Static & $\begin{aligned}
     &P^{}_{vac}(\Omega)=\frac{\lambda^{2}\hbar_f}{2 a \Omega^{2}\,\hbar_{d}\,}\frac{1}{e^{\frac{2\pi\omega }{a}}-1}  \\
     \label{p-d-1s}
&    P^{}_{vac}(-\Omega)  =\frac{\lambda^{2}\hbar_f}{2 a \Omega^{2}\,\hbar_{d}\,}\frac{1}{1-e^{-\frac{2\pi\omega }{a}}}  \end{aligned}$&   
    $\begin{aligned}
    &    {P}^{}_{\alpha}(\pm\Omega)= \frac{\lambda^{2}\alpha_{k}^{2}\hbar_f}{4 a\Omega^{2}\,\hbar_{d}\,\sinh\frac{\pi\omega}{a}}\;   \Big|
   e^{\frac{\pi\omega }{2a}} e^{i\phi}
       - e^{-\frac{\pi\omega }{2a}}e^{-i\phi} \Big|^{2}
    \end{aligned}$\\\cline{1-3}
    
 \hline

\end{tabular}

     }      
\end{table}
\end{center}
\end{widetext}

\section{Atomic transitions in $(3+1)$-dimensional spacetime}\label{s4}
In the earlier analysis, we considered $(1+1)$ dimensional spacetime to study the detector transition probabilities in two scenarios: $(i)$ the atom is accelerating with uniform acceleration, while the mirror is static, and $ (ii)$ the mirror is uniformly accelerating, while the atom is static in Minkowski spacetime. Now we will study the same in $(3+1)$-dimensional spacetime.

\subsection{The atom is uniformly accelerating} 
In this scenario, we consider the mirror is kept at $z=z_{0}$ and extended in the $x$-$y$ plane. Here, the mode functions of a massless real scalar field in the Minkowski spacetime are known as $ u_{k}^{M}(x)=e^{-i \omega t+i\vec{k}_{\perp}\cdot\vec{x}_{\perp}}(e^{i k_{z}(z-z_{0})}-e^{-i k_{z}(z-z_{0})})/\sqrt{(2\pi)^3\,2\omega}\,.$
    Here we take $k_z=\omega\cos\theta$ and $k_{\perp}=(k_x,\,k_y)$. $\theta$ is the polar angle between $\vec{k}$ and the direction of acceleration. In this case, the trajectory of the detector can be taken similar to Eq. (\ref{xtD}). Here we consider the acceleration along the $z$-direction and keep the other coordinates ($x,\,y$) to be fixed.

The Minkowski modes can be expressed along the worldline of the accelerated atom using this coordinate transformation as
\begin{eqnarray}\label{u3M-wmD}
    &&u^{M}_k=\frac{e^{ik_{\perp}\cdot x_{\perp}}}{\sqrt{16\pi^3\omega}}[F\,e^{-i(Ae^{a\tau}-Be^{-a\tau})}-F^{\star}\,e^{-i(Be^{a\tau}-Ae^{-a\tau})}]\,.~~
\end{eqnarray}
where we defined \begin{equation}\label{AB}
  A= \frac{\omega(1-\cos\theta)}{2a};~~~B=\frac{\omega(1+\cos\theta)}{2a};~~~F=e^{-i\omega z_0\cos\theta }\,.
    \end{equation}
 Using this field mode in Eq. (\ref{u3M-wmD}) along with Eq. (\ref{p-vac}), we can calculate the vacuum contribution in detector transition probabilities. However to obtain an analytic expression, we consider a condition $\frac{\omega}{a}<<1$ for which the latter calculation will be done. In this regime, the vacuum parts have been derived in \cite{Kumawat:2024kul}. The results are
\begin{eqnarray}
    &&{P}_{vac}(\Omega)
        = \frac{\hbar_f\lambda^2}{2\pi^2a\omega\Omega\,\hbar_{d}}
        \frac{\big|\sin{(k_{z}z_{0}-\psi_{1})}+\sin{(k_{z}z_{0}-\psi_{2})}\big|^2}{e^{\frac{2\pi\Omega}{a}}-1}~;
        \label{P-e-wm3-2}
        \nonumber
        \\
        \\
        &&{P}_{vac}(-\Omega) = \frac{\hbar_f\lambda^2}{2\pi^2a\omega\Omega\,\hbar_{d}}  \frac{  \big|\sin{(k_{z}z_{0}+\psi_{1})}+\sin{(k_{z}z_{0}+\psi_{2})}\big|^2}{1-e^{-\frac{2\pi\Omega}{a}}},
    \label{P-d-wm3-2}
    \nonumber
        \\
\end{eqnarray}
where we defined $\psi_{j}=\text{Arg}[\kappa_{j}]$ with $j=1,2$ . The expressions of $\kappa_{1}$ and $\kappa_{2}$ are given by $ \kappa_1=\big(\frac{\omega+k_{z}}{2a}\big)^{-\frac{i \Omega }{a}}\Gamma\big(\frac{i\Omega }{a}\big)$ and $\kappa_2=\big(\frac{\omega-k_{z}}{2a}\big)^{\frac{i \Omega }{a}}\Gamma\big(-\frac{i\Omega }{a}\big)$.  The vacuum contributions in excitation and de-excitation probabilities follow the relation $P_{vac}(\Omega)/P_{vac}(-\Omega)=e^{-\frac{2\pi\Omega}{a}}$ if $z_0=0$.

Now using Eq. (\ref{p-c}) and (\ref{u3M-wmD}), we can evaluate the non-vacuum contributions in the detector's excitation probability as 
\begin{eqnarray}
    &&P_{\alpha}(\Omega)=\frac{\lambda^2\alpha_k^2\,\hbar_{f}}{16\pi^3\omega\,\hbar_{d}}\Big|
    \int_{-\infty}^{\infty}d\tau\,e^{-i\Omega\tau}
    \nonumber
    \\
    &\times&\Big\{e^{ik_{\perp}\cdot x_{\perp}}\Big[F\,e^{-i(Ae^{a\tau}-Be^{-a\tau})}-F^{\star}\,e^{-i(Be^{a\tau}-Ae^{-a\tau})} \Big]\nonumber\\
   &+&e^{-ik_{\perp}\cdot x_{\perp}} \Big[F^{\star}\,e^{i(Ae^{a\tau}-Be^{-a\tau})}-F\,e^{i(Be^{a\tau}-Ae^{-a\tau})}\Big]
    \Big\}\Big|^2\nonumber\\
    &=&\frac{\lambda^2\alpha_k^2\,\hbar_{f}}{16\pi^3\omega\,\hbar_{d}}\Big|
    \int_{0}^{\infty}\frac{dy}{ay}\,y^{-\frac{i\Omega}{a}}\Big\{
    e^{ik_{\perp}\cdot x_{\perp}}\Big[F\,e^{-(e^{\frac{i\pi}{2}}Ay+e^{\frac{-i\pi}{2}}\frac{B}{y})}
    \nonumber
    \\
    &-&F^{\star}\,e^{-(e^{\frac{i\pi}{2}}By+e^{\frac{-i\pi}{2}}\frac{A}{y})} \Big]\nonumber\\
   &+& e^{-ik_{\perp}\cdot x_{\perp}} \Big[F^{\star}\,e^{-(e^{\frac{-i\pi}{2}}Ay+e^{\frac{i\pi}{2}}\frac{B}{y})}-F\,e^{-(e^{\frac{-i\pi}{2}}By+e^{\frac{i\pi}{2}}\frac{A}{y})}\Big]
    \Big\}\Big|^2~.
    \nonumber
    \\
    \end{eqnarray}
Here, we have used the change of variable $e^{a\tau}=y$. We will evaluate the above integral using the following integral expression of the modified Bessel function \cite{Gradshteyn}:
\begin{equation}\label{besselK}
    K_{\mu}(xz)=\frac{z^{\mu}}{2}\int_0^{\infty}dt\,t^{-\mu-1}\exp\Big[-\frac{x}{2}\Big(t+\frac{z^2}{t}\Big)\Big]\,.
\end{equation}
We obtain

\begin{eqnarray}\label{P-a-e-3a-A}
    &&P_{\alpha}(\Omega)
    =\frac{\lambda^2\alpha_k^2\,\hbar_{f}}{4\pi^3a^2\omega\,\hbar_{d}}   
    \Big|{K}_{\frac{i\Omega}{a}}\Big(\frac{\omega}{a}\sin{\theta}\Big)\Big[
    e^{-ik_{\perp}\cdot x_{\perp}+\frac{\pi\Omega}{2a}} 
   \Big( \Big(\frac{A}{B}\Big)^{\frac{i\Omega}{2a}} F^{\star}
   \nonumber
   \\
   &-&\Big(\frac{A}{B}\Big)^{-\frac{i\Omega}{2a}}F\Big)
+e^{ik_{\perp}\cdot x_{\perp}-\frac{\pi\Omega}{2a}}\Big(\Big(\frac{A}{B}\Big)^{\frac{i\Omega}{2a}}F-
    \Big(\frac{A}{B}\Big)^{-\frac{i\Omega}{2a}}F^{\star}\Big)\Big]\Big|^2.~~~~~~
\end{eqnarray}
In the large-acceleration limit ($\omega/a \ll 1$), the above expression can be simplified as (see Appendix~(\ref{Appendix_WM_3a}) for details)

\begin{equation}\label{P-a-e-3a}
    \begin{aligned}
     &  P_{\alpha}(\Omega)  =\frac{\lambda^2\alpha_k^2\hbar_f}{4\pi^2a\omega\Omega\,\hbar_{d}\sinh\frac{\pi\Omega}{a}}   \\&
     ~~~~~~~~~~~~\times\Big|
    e^{-ik_{\perp}\cdot x_{\perp}+\frac{\pi\Omega}{2a}} 
   \Big( \sin{(k_{z}z_{0}+\psi_{1})}+
   \sin{(k_{z}z_{0}+\psi_{2})} \Big)\\&
  ~~~~~~~~~~~~~~~ 
    -e^{ik_{\perp}\cdot x_{\perp}-\frac{\pi\Omega}{2a}}\Big(\sin{(k_{z}z_{0}-\psi_{1})}+\sin{(k_{z}z_{0}-\psi_{2})}\Big)\Big|^2\,.
    \end{aligned}
\end{equation}

\subsection{The atom is static }   
In this setup, we keep the atom static at $z=z_{0}$ in the Minkowski spacetime, and the mirror is uniformly accelerating along the $z$ direction. In the Rindler frame, the mirror remains fixed at $\xi=0$. The field mode solution in the mirror frame is given by (see Ref.~\cite{Kumawat:2024kul})
\begin{equation}\label{u3R-wmD}
    u^{R}_{k} =\sqrt{\frac{\sinh{\frac{\pi\omega}{a}}}{4\pi^4a}}\,~e^{i\,\vec{k}_{\perp}\cdot\vec{x}_{\perp}-i\,\omega\,\eta}[{K}_{\frac{i\omega}{a}}\bigg(\frac{k_{\perp}e^{a\,\xi}}{a}\bigg)-{K}_{\frac{i\omega}{a}}\bigg(\frac{k_{\perp}e^{-a\,\xi}}{a}\bigg)]~.
\end{equation}
In the large-acceleration limit and for $k_{\perp}e^{a\xi}/a\ll1$, using Eq.~(\ref{besselKasym}), the mode function can be rewritten as
\begin{equation}
    u_k^R=U(e^{-i\omega(\eta-\xi)}-e^{-i\omega(\eta+\xi)})+V(e^{-i\omega(\eta+\xi)}-e^{-i\omega(\eta-\xi)})
\end{equation}
where $U$ and $V$ are defined by
\begin{equation}\label{UV}
     \begin{aligned}
& U=\sqrt{\frac{\sinh{\frac{\pi\omega}{ a}}}{16\pi^4a}} e^{i\,\vec{k}_{\perp}\cdot\vec{x}_{\perp}}\Gamma\biggl(-\frac{i\omega}{a}\biggl)\,\bigg(\frac{k_{\perp}}{2a}\bigg)^{\frac{i\omega}{a}}\,;\\
&V=\sqrt{\frac{\sinh{\frac{\pi\omega}{ a}}}{16\pi^4a}} e^{i\,\vec{k}_{\perp}\cdot\vec{x}_{\perp}}\Gamma\biggl(\frac{i\omega}{a}\biggl)\,\bigg(\frac{k_{\perp}}{2a}\bigg)^{-\frac{i\omega}{a}}\,.
     \end{aligned}
 \end{equation}
Finally, the mode function is expressed in terms of the Minkowski coordinates and takes the form
\begin{equation}\label{u3Rm}
    u^{R}_{k}=(U-V)[\theta(z_0-t)\,\big(a(z_0-t)\big)^{\frac{i\omega}{a}} -\theta(z_0+t)\,\big(a(z_0+t)\big)^{-\frac{i\omega}{a}}] \,.
\end{equation}
The above mode can be extended over the entire proper time of the detector, as discussed in Sec.~(\ref{Sec1plus1static}).

   The detector excitation and de-excitation probabilities from the field vacuum can be evaluated using Eq. (\ref{p-vac}) and (\ref{u3Rm}), and are known as \cite{Kumawat:2024kul}
 \begin{eqnarray}\label{P-e-wm3-s}
     &&P_{vac}(\Omega)=\frac{2\hbar_f\lambda^2\sin^2\varphi_4}{\pi^2a\Omega^2\,\hbar_{d}}\frac{\sin^2(\Omega z_0+\phi)}{e^{\frac{2\pi\omega}{a}}-1}\,;
     \\
\label{P-d-wm3-s}     &&P_{vac}(-\Omega)=\frac{2\hbar_f\lambda^2\sin^2\varphi_4}{\pi^2a\Omega^2\,\hbar_{d}}\frac{\sin^2(\Omega z_0-\phi)}{1-e^{-\frac{2\pi\omega}{a}}}\,,
 \end{eqnarray}
where  $\phi=\text{Arg}[\bigl(\frac{\Omega}{a}\bigl)^{\frac{i\omega}{a}}\;\Gamma\bigl(-\frac{i\omega }{a}\bigl)]$ and $\varphi_4=\text{Arg}[\bigl(\frac{k_{\perp}}{2a}\bigl)^{\frac{i\omega}{a}}\;\Gamma\bigl(-\frac{i\omega }{a}\bigl)]$. 
Now, from Eq. (\ref{p-c}) and (\ref{u3Rm}), one can calculate non-vacuum contributions in the transition probability and obtain it as (see Appendix~(\ref{Appendix_WM_3s}) for details)

\begin{equation}\label{P-a-e-3s}
    \begin{aligned}
       & P_{\alpha}(\Omega)
=\frac{\lambda^2\alpha_k^2\hbar_f\sin^2\varphi_4}{\pi^2a\Omega^2\,\hbar_{d}\,\sinh\frac{\pi\omega}{a}}\Big|e^{-\frac{\pi\omega}{a}+ik_{\perp}\cdot x_{\perp}}\sin{(\Omega z_0+\phi)}\\&
~~~~~~~~~~~~~~~~~~~~~~~~~~~~~~~~~+e^{\frac{\pi\omega}{a}-ik_{\perp}\cdot x_{\perp}}\sin{(\Omega z_0-\phi)}\Big|^2\,.
    \end{aligned}
\end{equation}

The corresponding expressions in the absence of boundary have been calculated in Appendix \ref{Appendix_NM_3}. We summarize all the results in tables \ref{T3} and \ref{T4}.

\begin{widetext}
\begin{center} 
\begin{table}[h!]
\caption{The transition probability of the atom with a mirror in the $(3+1)$ dimensions }
     \label{T3}
 
 {
    \centering
    \begin{tabular}{ |c|c|c|  }
 \hline

 \multicolumn{3}{|c|}{Transition Probability in $(3+1)$ Dimensions with Mirror $( P_{\omega}(\Omega)= P_{vac}(\Omega)+ P_{\alpha}(\Omega))$} \\
 \hline    Atom accelerated/static & Vacuum contribution  & Non-vacuum contribution \\
 \hline  Accelerated & $\begin{aligned}
       &    {P}_{vac}(\Omega)
        = \frac{\hbar_f\lambda^2}{2\pi^2a\omega\Omega\,\hbar_{d}\,}
        \frac{\Big|\sin{(k_{z}z_{0}-\psi_{1})}+\sin{(k_{z}z_{0}-\psi_{2})}\Big|^2}{e^{\frac{2\pi\Omega}{a}}-1}\\
    &  {P}_{vac}(-\Omega) = \frac{\hbar_f\lambda^2}{2\pi^2a\omega\Omega\,\hbar_{d}\,}  \frac{  \Big|\sin{(k_{z}z_{0}+\psi_{1})}+\sin{(k_{z}z_{0}+\psi_{2})}\Big|^2}{1-e^{-\frac{2\pi\Omega}{a}}}
        \end{aligned}$&    
        $\begin{aligned}
         &{P}^{}_{\alpha}(\pm\Omega)
          =\frac{\lambda^2\alpha_k^2\hbar_f}{4\pi^2a\omega\Omega\,\hbar_{d}\sinh\frac{\pi\Omega}{a}}   \\&
     ~~~~~~~~~~~~\times\Big|
    e^{-ik_{\perp}\cdot x_{\perp}+\frac{\pi\Omega}{2a}} 
   \Big( \sin{(k_{z}z_{0}+\psi_{1})}+
   \sin{(k_{z}z_{0}+\psi_{2})} \Big)\\&
  ~~~~~~~~~~~~~~~ 
    -e^{ik_{\perp}\cdot x_{\perp}-\frac{\pi\Omega}{2a}}\Big(\sin{(k_{z}z_{0}-\psi_{1})}+\sin{(k_{z}z_{0}-\psi_{2})}\Big)\Big|^2
         \end{aligned}$\\ \cline{1-3}
    Static & $\begin{aligned}
   &P_{vac}(\Omega)=\frac{2\hbar_f\lambda^2\sin^2\varphi_4}{\pi^2a\Omega^2\,\hbar_{d}\,}\frac{\sin^2(\Omega z_0+\phi)}{e^{\frac{2\pi\omega}{a}}-1}\\
   &P_{vac}(-\Omega)=\frac{2\hbar_f\lambda^2\sin^2\varphi_4}{\pi^2a\Omega^2\,\hbar_{d}\,}\frac{\sin^2(\Omega z_0-\phi)}{1-e^{-\frac{2\pi\omega}{a}}}
     \end{aligned}$&   
    $\begin{aligned}
        &{P}^{}_{\alpha}(\pm\Omega)=
      \frac{\lambda^2\alpha_k^2\hbar_f\sin^2\varphi_4}{\pi^2a\Omega^2\,\hbar_{d}\,\sinh\frac{\pi\omega}{a}}\Big|e^{-\frac{\pi\omega}{a}+ik_{\perp}\cdot x_{\perp}}\sin{(\Omega z_0+\phi)}\\&
~~~~~~~~~~~~~~~~~~~~~~~~~~~~~~~~~~~~~~~~+e^{\frac{\pi\omega}{a}-ik_{\perp}\cdot x_{\perp}}\sin{(\Omega z_0-\phi)}\Big|^2
    \end{aligned}$\\\cline{1-3}
    
 \hline
\end{tabular}
    
        }
\end{table}

 \begin{table}[h!]
  \caption{The transition probability of the atom without a mirror in the $(3+1)$ dimensions. In the following expressions, $\varphi_{3}$ and $\varphi_{ex}$ are defined as  $\varphi_3=Arg\left[\left(\omega\sin{\theta}/2a\right)^{i \Omega /a}\Gamma\left(- i\Omega/a\right)\right]$ and $\varphi_{ex}=Arg[(k_{\perp}/2\Omega)^{-i\omega/a}]$.}
     \label{T4}
 {
    \centering
    \begin{tabular}{ |c|c|c|  }
 \hline

 \multicolumn{3}{|c|}{Transition Probability without boundary in $(3+1)$ Dimensions $( P_{\omega}(\Omega)= P_{vac}(\Omega)+ P_{\alpha}(\Omega))$} \\
 \hline    Atom accelerated/static & Vacuum contribution  & Non-vacuum contribution \\
 \hline  Accelerated & $\begin{aligned}&{P}_{vac}(\Omega)
        = \frac{\hbar_f\lambda^2}{2\pi^2\,a\,\omega\,\Omega\,\hbar_{d}\,}\frac{\cos^2\varphi_3}{e^{\frac{2\pi\Omega}{a}}-1}\\
        
 &{P}_{vac}(-\Omega)
        = \frac{\hbar_f\lambda^2}{2\pi^2\,a\,\omega\,\Omega\,\hbar_{d}\,}\frac{\cos^2\varphi_3}{1-e^{-\frac{2\pi\Omega}{a}}}
        \end{aligned}$&    
        $\begin{aligned}
         &{P}^{}_{\alpha}(\pm\Omega)
         =\frac{\lambda^2\alpha_{k}^{2}\hbar_f \cos^2\varphi_3}{4\pi^{2}\omega a\Omega\,\hbar_{d}\,\sinh{\frac{\pi\Omega}{a}}}
    \Big|e^{i\vec{k}_{\perp}.\vec{x}_{\perp}}e^{\frac{\pi  \Omega }{2 a}} +e^{-i\vec{k}_{\perp}.\vec{x}_{\perp}}e^{-\frac{\pi  \Omega }{2 a}}\Big|^{2}
         \end{aligned}$\\ \cline{1-3}
    Static & $\begin{aligned}
    &{P}_{vac}(\Omega)
        = \frac{\hbar_f\lambda^2}{2\pi^2\,a\,\Omega^2\,\hbar_{d}\,}\,\frac{\cos^2(\Omega z_0+\varphi_{ex})}{e^{\frac{2\pi\omega}{a}}-1}~;\\
       
     &  {P}_{vac}(-\Omega) = \frac{\hbar_f\lambda^2}{2\pi^2\,a\,\Omega^2\,\hbar_{d}\,}\,\frac{\cos^2(\Omega z_0-\varphi_{ex})}{1-e^{-\frac{2\pi\omega}{a}}}
     \end{aligned}$&   
    $\begin{aligned}
        &{P}^{}_{\alpha}(\pm\Omega)
        =\frac{\lambda^2\alpha_{k}^{2}\hbar_f}{4\pi^2a\Omega^2\,\hbar_{d}\,\sinh{(\frac{\pi\omega}{a})}}
\big|-e^{-\frac{\pi\omega }{2a}} e^{i\,\vec{k}_{\perp}.\vec{x}_{\perp}}\cos{(\Omega z_0+\varphi_{ex})}\\
&~~~~~~~~~~~~~~~~~~~~~~~~~~~~~~~~~~~~~~~~~~~~~~+e^{\frac{\pi\omega }{2a}} e^{-i\,\vec{k}_{\perp}.\vec{x}_{\perp}}\cos{(\Omega z_0-\varphi_{ex})}\big|^{2}
    \end{aligned}$\\\cline{1-3}
    
 \hline
\end{tabular}
   
        }
\end{table}
\end{center}
\end{widetext}

\section{``Semi''-classical limits}\label{s5}
Note that in $(1+1)$-dimensions, the transition probabilities from vacuum in either accelerated and static situations are identical when one takes $\Omega=\omega$. This is already known in literature (see \cite{PhysRevLett.121.071301}). For $(3+1)$-dimensions, the excitation to de-excitation ratio, i.e., $P_{vac}(\Omega)/P_{vac}(-\Omega)$, with $z_0=0$, shows similarity between the two motions of the atom once $\Omega=\omega$ is taken \cite{Barman:2024dql,Kumawat:2024kul}. However, the presence of $P_\alpha$’s broke that symmetry. 
In our analysis, we considered a coherent field state, which is known to be the closest approximation to a classical field. We now analyse the transition probabilities in two different limits: (A) the field is classical, and (B) both the field and the atom are in a classical regime.

The meaning of the two semi-classical limits $\hbar_d\to 0$ and $\hbar_f\to 0$ can be understood in the following example. The physically relevant quantities are dimensionless ratios of the form $S/\hbar$, where $S$ is the action. In our setup, the corresponding actions are $S_d = \int d\tau \, \mathcal{L}_d$ and $S_f = \int d^4x \sqrt{-g}\, \mathcal{L}_f$, where we do not assume any explicit form for the detector Lagrangian, while for the field we consider a massless scalar with $\mathcal{L}_f = \partial_a \phi \, \partial^a \phi$.
The $\hbar \to 0$ limit is understood in the standard semiclassical sense, where the wavefunction is taken in the form $\psi \sim \exp\left(\frac{i}{\hbar} S\right)$ and $S$ is expanded as $S=S_{\text{cl}} + \hbar S_1 + \hbar^2 S_2+\dots$. In the limit $\hbar \to 0$, the dominant contribution is given by the classical action $S_{\text{cl}}$, so that $\psi \sim \exp\left(\frac{i}{\hbar} S_{\text{cl}}\right)$, with quantum corrections suppressed by powers of $\hbar$.
Thus, taking $\hbar_d \to 0$ or $\hbar_f \to 0$ corresponds to the semiclassical limits $S_d/\hbar_d \gg 1$ and $S_f/\hbar_f \gg 1$, which makes the interpretation of these limits explicit.

\subsection{``Semiclassical'' field}
In order to obtain the response of the (semi)classical field on the quantum atom, one needs to take the limit $\hbar_f\to0$ by keeping $\alpha_k^2\hbar_f$ finite. Physically, this means that quantum vacuum fluctuations of the field vanish as $\hbar_f\to0$, while the field amplitude and energy remain finite because the combination $\alpha_k^{2}\hbar_f$ is kept fixed.  If $\hbar_f$ were taken to zero with $\alpha_k$ fixed, the semiclassical field would disappear, whereas keeping $\alpha_k^{2}\hbar_f$ finite yields a nonzero classical wave that acts as a deterministic background for the quantum detector. This can be realised in the following way.
The random classical scalar field in $(n+1)$ dimensions is defined (see  Eq.~(30) of Ref.~\cite{PhysRevD.21.2137}) as
\begin{equation}\label{clasical field}
    \phi(\vec{r},t)
      = \int d^{n}k\, \frac{1}{2}f(\omega)
      \left[
      \mu_k \, e^{-i(\omega t - \vec{k}\cdot\vec{r})}
      + \mu^{*}_k \, e^{\,i(\omega t - \vec{k}\cdot\vec{r})}
      \right],
\end{equation}
where $\mu_k=e^{-i\theta(\vec{k})}$ carries a random phase $\theta(\vec{k})$.
On the other hand, the free quantum scalar field in $(n+1)$ dimensions, written in terms of the annihilation and creation operators (see  Eq.~(34) of Ref.~\cite{PhysRevD.21.2137}), is given by
\begin{equation}\label{Quantum field}
   \hat{\phi}(\vec{r},t)
      = \int \frac{d^{n}k}{(2\pi)^{n/2}}  \sqrt{\frac{\hbar\,c^{2}}{2\omega}}
      \left[
      \hat{a}(\vec{k}) \, e^{-i(\omega t - \vec{k}\cdot\vec{r})}
      + \hat{a}^{\dagger}(\vec{k}) \, e^{\,i(\omega t - \vec{k}\cdot\vec{r})}
      \right],
\end{equation}
 By comparing the ensemble average of the classical field (\ref{clasical field}) with the expectation value of the quantum field (\ref{Quantum field}) in a coherent state, one finds that the classical mode amplitude $f(\omega)$ and the coherent-state amplitude $\alpha_k$ are related through $\beta f^{2} (\omega)\omega=\alpha_{k}^{2}\hbar$, where $\beta$ is a constant given by  $\beta=(2\pi)^{n}/2\,c^{2}$. In particular, for $(1+1)$ dimensions one has $\beta=\pi/c^{2}$, whereas for $(3+1)$ dimensions $4\pi^{3}/c^{2}$.

In this (semi)classical-field limit, the vacuum contribution to the transition probabilities vanishes, i.e., $P_{{vac}}(\pm \Omega)=0$. This is expected, since it has no classical counterpart. However, the $P_\alpha$ contribution remains non-vanishing in all cases, and this quantity does not reflect the symmetry in the relative classical motions between the two frames. 
Therefore, the total response of the quantum atom to the classical field does not bear symmetry between the relative classical motions, except at the level of the vacuum excitation-to-de-excitation probability ratio. 
Such observation indicates two implications -- (i) The symmetry in the relative motions of the frame is more prone to the field vacuum state, rather than other field states. (ii) The classicality of the field responds to a quantum system in a frame-dependent way.

\subsection{Semi-classicality of both field and detector}{\label{Sec:Climit_F&D}}
We just observed that the quantum detector's response to the (semi)classical field does not respect the symmetry in the two scenarios. However, since the aforesaid symmetry is known to be valid in the classical trajectories, it would be interesting to investigate whether the transition probabilities can reflect such symmetry in the semiclassical limit of both field and detector. In the above discussion, we examined the semiclassical limits of the fields ($\hbar_f\to 0$) while keeping $\alpha_k^2\hbar_f = \beta\omega f^{2}$ fixed. Now, on top of this, we will take the $\hbar_d\to0$ limit. The later limit will provide the transition probabilities of the atom when the behaviour of both field and atom is very close to a classical one. To impose the latter limit, we consider a simple situation where $\omega=\Omega$ (the resonance condition) to simplify the analysis and provide insight through an analytical calculation that yields closed-form expressions. Furthermore, in doing so, we will use $\omega=\Omega=\frac{\Delta E}{\hbar_d}$, where $\Delta E$ is the energy gap between two quantum energy levels of the atom and keep $\Delta E$ fixed while taking the limit $\hbar_d\to 0$. 

Under these assumptions, the effective-field contribution to the transition probability in $(1+1)$ dimensions simplifies considerably. For an accelerated atom in the presence of a static mirror, we find that $\varphi_{1}=0$, since $\Gamma(\pm i\infty)=0$. Consequently, Eq.~(\ref{P-a-e-1a}) reduces to
\begin{equation}
\label{eq:P-a-e-1a-Climit-F&D}
P_{\alpha}
=\frac{2\lambda^{2}\beta f^{2}}{\Delta E\,a}
\lim_{\hbar_d\to0}\sin^2\!\left(\frac{\Delta E z_{0}}{\hbar_d}\right).
\end{equation}
Similarly, for a static atom with an accelerated mirror, we find that $\varphi_{2}=0$, since $\Gamma(\pm i\infty)=0$. Consequently, Eq.~(\ref{P-a-e-1s}) reduces to that of the accelerated atom case (see Eq. (\ref{eq:P-a-e-1a-Climit-F&D})).
The final expressions are summarised in Table~(\ref{Table:CL_WM1}). For completeness, the corresponding results in the absence of the mirror are also listed in Table~(\ref{Table:CL_NM1}), with detailed derivations given in Appendix~(\ref{AppB1}). Throughout this limiting procedure in $(1+1)$ dimensions, the acceleration $a$ is kept finite.

\begin{widetext}
    \begin{center}
 \begin{table}[h!]
  \caption{The transition probability of the atom with a mirror in the $(1+1)$ dimensions, obtained in the semiclassical limits of both the field and the
    detector. }
    \label{Table:CL_WM1} 
 {
    \centering
    \begin{tabular}{ |c|c|c|  }
 \hline
\multicolumn{3}{|c|}{Transition Probability in $(1+1)$ Dimensions with Mirror $( P_{\omega}(\Omega)= P_{vac}(\Omega)+ P_{\alpha}(\Omega))$} \\
 \hline    Atom accelerated/static & Vacuum contribution  & Non-vacuum contribution \\
 \hline  Accelerated & 
    $\begin{aligned}
    &{P}^{}_{vac}(\pm\Omega) 
    = 0
    \end{aligned}$&    
        $\begin{aligned}
         &{P}^{}_{\alpha}(\pm\Omega)=\frac{2\lambda^{2}\beta f^{2}}{\Delta E\,a}
\left|\lim_{\hbar_d\to0}\sin\!\left(\frac{\Delta E z_{0}}{\hbar_d}\right)\right|^{2}.
         \end{aligned}$\\ \cline{1-3}
    Static & $\begin{aligned}
    &P_{vac}(\pm\Omega)= 0 
    \end{aligned}$&   
    $\begin{aligned}
        &{P}^{}_{\alpha}(\pm\Omega) =\frac{2\lambda^{2}\beta f^{2}}{\Delta E\,a}
\left|\lim_{\hbar_d\to0}\sin\!\left(\frac{\Delta E z_{0}}{\hbar_d}\right)\right|^{2}.
    \end{aligned}$\\\cline{1-3}
\hline
\end{tabular}
}
\end{table}
\begin{table}[h!] 
  \caption{The transition probability of the atom without a mirror in the $(1+1)$ dimensions in the semiclassical limit of both the field and the detector.}
     \label{Table:CL_NM1}
 {
    \centering
    \begin{tabular}{|c|c|c|  }
 \hline
\multicolumn{3}{|c|}{Transition Probability in $(1+1)$ Dimensions $( P_{\omega}(\Omega)= P_{vac}(\Omega)+ P_{\alpha}(\Omega))$} \\
 \hline    Atom accelerated/static & Vacuum contribution  & Non-vacuum contribution \\
 \hline  Accelerated & $\begin{aligned}
 &P_{vac}(\pm\Omega)
  =0 
   \end{aligned}$&    
        $\begin{aligned}
         {P}^{}_{\alpha}(\pm\Omega)&=\frac{\lambda^{2}\beta f^{2}}{2a\,\Delta E}.
         \end{aligned}$\\ \cline{1-3}
    Static & $\begin{aligned}
     &P^{}_{vac}(\pm\Omega)=0   \end{aligned}$&   
    $\begin{aligned}
    &    {P}^{}_{\alpha}(\pm\Omega)= \frac{\lambda^{2}\beta f^{2}}{2a\,\Delta E}.
    \end{aligned}$\\\cline{1-3}
    \hline
\end{tabular}
 } 
\end{table}
\end{center}
\end{widetext}

We observe that within the mentioned restrictions and in the fully classical regime, the transition probabilities in $(1+1)$-dimensions of accelerated and static atoms coincide, both in the presence and in the absence of the mirror. This symmetry is absent when only the field is classical while the detector remains quantum, and it is restored only when both the system (the atom) and the environment (the field) are taken to their semiclassical limits.

\begin{figure}[h!]
    \centering
    \includegraphics[width=0.7\linewidth]{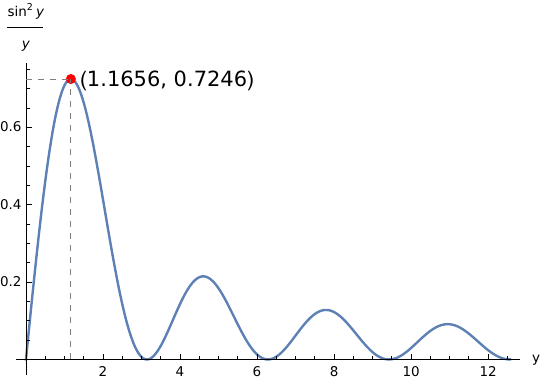}
    \caption{$Y= \sin^2y/y$ vs $y=(\Delta E z_0)/\hbar$ graph.}
    \label{fig:sin2y_over_y}
\end{figure}

In order to give the meaning of the expression Eq.~(\ref{eq:P-a-e-1a-Climit-F&D}), we rewrite the same as
\begin{eqnarray}
P_{\alpha}(\Omega)
            &=& \frac{2\lambda^{2}\beta f^{2}}{a}\lim_{\hbar_{d}\to0}\Big(\frac{z_{0}}{\hbar_{d}}\Big)\underbrace{\frac{1}{\frac{\Delta E z_{0}}{\hbar_{d}}}\sin^{2}\!\left(\frac{\Delta E z_{0}}{\hbar_{d}}\right)}_{Y}~.
\end{eqnarray}
Introducing the dimensionless variable $y=(\Delta E\,z_{0})/\hbar_{d}$, the quantity inside the under-brace becomes $Y=\sin^{2}y/y$. The behaviour of this function is shown in Fig.~\ref{fig:sin2y_over_y}, where one observes a dominant central peak followed by damped oscillations. 
The first zero occurs at $y=\pi$, so that the effective width around the highest peak is of order $\pi$. This corresponds to
\begin{equation}
y=\pi=\frac{\Delta E\,z_{0}}{\hbar_{d}}
\quad\Rightarrow\quad
\Delta E=\frac{\pi\,\hbar_{d}}{z_{0}}~.
\label{E}
\end{equation}
The maximum of the function $\sin^{2}y/y$ is a pure number.  Numerically, the peak value is $0.7246$ at $y=1.1656$.  Therefore, the effective-field contribution to the transition probability at its maximum is
\begin{equation}
P_{\alpha}(\Omega)
= \frac{2\lambda^{2}\beta f^{2}}{a}
\lim_{\hbar_{d}\to0}\frac{z_{0}}{\hbar_{d}}\,(0.7246)\, .
\end{equation}
Hence, in the limit $\hbar_{d}\to 0$, the energy spacing $\Delta E$ between the atomic levels decreases, and the spectrum becomes effectively continuous (see Eq. (\ref{E})). This is expected as the quantum detector becomes close to classical one. 
On the other hand, the height of the dominant peak of $P_{\alpha}$ grows as $\hbar_{d}\to0$, indicating that the transition probability becomes increasingly sharply peaked in the classical-detector limit.

In $(3+1)$ dimensions, the situation is more subtle. In order to obtain analytic expressions for the transition probabilities, the calculations were performed in the regime $\frac{\omega}{a}\ll1$, which, upon imposing the resonance condition, becomes $\frac{\Delta E}{\hbar_d a}\ll1$. Consequently, the limit $\hbar_d \to 0$ cannot be taken independently in terms involving $\Delta E/(\hbar_d a)$, since the condition $\Delta E \ll \hbar_d a$ must be satisfied. This corresponds to a small energy gap $\Delta E$, yielding a quasi-continuous spectrum and an approximately classical detector. Therefore, the simplified results in the regime $\Delta E/(\hbar_d a)\ll1$ correspond to a situation in which both the field and the detector are close to their classical counterparts.
We obtain the transition probability of the uniformly accelerating detector in the presence of the static mirror as
\begin{eqnarray}\label{P-a-e-3a-CL_F&D}
  P_{\alpha}(\Omega) &\simeq&\frac{4\,\lambda^{2}\beta f^{2}}{\pi^{3}\,\hbar_{d}\,a^{2}\,}\Bigg[\sin^{2}\!\left(\frac{\Delta E}{\hbar_d} z_0\cos\theta\right)\,\sin^{2}(\vec{k}_{\perp}\!\cdot\!\vec{x}_{\perp}) \,\nonumber\\
  &&~~~~~~~~~~~~~~~~~~\times \left(\gamma+\ln\!\left[\frac{\Delta E}{2\hbar_d a}\sin\theta\right]\right)^{2}\Bigg]~.
\end{eqnarray}
While that for the static detector with the mirror is in uniform acceleration is given by
\begin{eqnarray}\label{P-a-e-3s-CL_F&D}
    P_{\alpha}(\Omega) &\simeq& \frac{4\,\lambda^2\beta \,f^{2}\hbar_{d}}{\pi^{3}\,\Delta E^{2}}\left[\cos^{2}\!\left(\frac{\Delta E}{\hbar_{d}} z_0\right)\,{\sin^{2}(\vec{k}_{\perp}\!\cdot\!\vec{x}_{\perp})}\right]. 
\end{eqnarray}
The intermediate steps are given in Appendix~(\ref{Appendix: WM_3}), and the results are summarised in Table~(\ref{Table:CL_WM3}). The corresponding expressions in the absence of the mirror are also provided in Table~(\ref{Table:CL_NM3}), with full derivations in Appendix~(\ref{Appendix: NM_3_Climit}).

\begin{widetext}

\begin{table}[H]
\caption{Transition probability of an atom in $(3+1)$ dimensions in the presence of a mirror, obtained in the classical-field limit and under the resonance condition $\omega=\Omega=\Delta E/\hbar_{d}$ in the large-acceleration regime $\Delta E/(\hbar_{d}a)\ll1$.}
     \label{Table:CL_WM3}
    \centering
    \begin{tabular}{ |c|c|c|  }
 \hline

 \multicolumn{3}{|c|}{Transition Probability in $(3+1)$ Dimensions with Mirror $( P_{\omega}(\Omega)= P_{vac}(\Omega)+ P_{\alpha}(\Omega))$} \\
 \hline    Atom accelerated/static & Vacuum contribution  & Non-vacuum contribution \\
 \hline  Accelerated & $\begin{aligned}
       &    {P}_{vac}(\pm\Omega)
        = 0
        \end{aligned}$&    
        $\begin{aligned}
         &{P}^{}_{\alpha}(\pm\Omega)
          \simeq \frac{4\,\lambda^{2}\beta f^{2}}{\pi^{3}\,\hbar_{d}\,a^{2}\,}\Bigg[\sin^{2}\!\left(\frac{\Delta E}{\hbar_d} z_0\cos\theta\right)\,\sin^{2}(\vec{k}_{\perp}\!\cdot\!\vec{x}_{\perp}) \left(\gamma+\ln\!\left[\frac{\Delta E}{2\hbar_d a}\sin\theta\right]\right)^{2}\Bigg]
         \end{aligned}$\\ \cline{1-3}
    Static & $\begin{aligned}
   &P_{vac}(\pm\Omega)=0
     \end{aligned}$&   
    $\begin{aligned}
        &{P}^{}_{\alpha}(\pm\Omega)\simeq  \frac{4\,\lambda^2\beta \,f^{2}\hbar_{d}}{\pi^{3}\,\Delta E^{2}}\left[\cos^{2}\!\left(\frac{\Delta E}{\hbar_{d}} z_0\right)\,{\sin^{2}(\vec{k}_{\perp}\!\cdot\!\vec{x}_{\perp})}\right]  
      
    \end{aligned}$\\\cline{1-3}
    
 \hline
\end{tabular}

\end{table}

 \begin{table}[H]
 \caption{Transition probability of an atom in $(3+1)$ dimensions without a mirror, obtained in the classical-field limit and under the resonance condition $\omega=\Omega=\Delta E/\hbar_{d}$ in the large-acceleration regime $\Delta E/(\hbar_{d}a)\ll1$.}
    \label{Table:CL_NM3}
    \centering
    \begin{tabular}{ |c|c|c|  }
 \hline

 \multicolumn{3}{|c|}{Transition Probability without boundary in $(3+1)$ Dimensions $( P_{\omega}(\Omega)= P_{vac}(\Omega)+ P_{\alpha}(\Omega))$} \\
 \hline    Atom accelerated/static & Vacuum contribution  & Non-vacuum contribution \\
 \hline  Accelerated & $\begin{aligned}&{P}_{vac}(\pm\Omega)
        = 0
        \end{aligned}$&    
        $\begin{aligned}
         &{P}^{}_{\alpha}(\pm\Omega)\simeq \frac{\lambda^{2}\beta f^{2}}{\pi^{3} \hbar_{d} a^{2}}\left[\left(\ln\!\left[\frac{\Delta E \sin\theta}{2\hbar_{d} a}\right]+\gamma\right)^{2}\, \cos^{2}(\vec{k}_{\perp}\!\cdot\!\vec{x}_{\perp})\right]
         \end{aligned}$\\ \cline{1-3}
    Static & $\begin{aligned}
    &{P}_{vac}(\pm\Omega)
        = 0
     \end{aligned}$&   
    $\begin{aligned}
        &{P}^{}_{\alpha}(\pm\Omega)\simeq \frac{\lambda^2 \beta f^2\,\hbar_{d}}{\pi^3  \Delta E^{2}}\, \sin^{2}(\vec{k}_{\perp}\!\cdot\!\vec{x}_{\perp})\, 
                
    \end{aligned}$\\\cline{1-3}
    
 \hline
\end{tabular}

\end{table}

\end{widetext}
We observe that while in $(1+1)$ dimensions the fully semiclassical limit leads to an exact matching between the transition probabilities of accelerated and static atoms, in $(3+1)$ dimensions the situation is more complicated; within our region of applicability, however, such a matching between the transition probabilities of accelerated and static atoms is no longer evident.

\section{Conclusion}\label{s6}
The accelerated frame observes a particle in the Minkowski vacuum, and conversely, the accelerated vacuum appears to a static observer to be filled with particles. Consequently, the relative non-inertial motion between the field (and so the vacuum) and the detector causes a transition in the detector's energy levels. 
Since the motion of the accelerated frame as seen by a Minkowski observer is analogous to the motion of the Minkowski observer as seen from the accelerated frame, one may ask whether this classical analogy is reflected in the corresponding quantum phenomena. This general viewpoint has been addressed by investigating the transition probabilities of a two-level quantum detector interacting with a single-mode scalar field. 
In one case, the detector is uniformly accelerating while the photon is defined in the Minkowski frame. In the other case, the detector is static in the Minkowski frame, and the photon is defined in an accelerated frame.  Using the photon's vacuum state, it has been shown that these two scenarios are equivalent when the frequencies of the detector and the photon mode are equal -- either at the level of transition probabilities \cite{PhysRevLett.121.071301} or at the level of the ratio between the excitation and de-excitation probabilities \cite{Barman:2024dql,Kumawat:2024kul}. 

Here, we addressed a similar question, but now the field is in a coherent state. The choice of this field state is inspired by the following facts. In these transitions, the modulation of the field mode frequency with respect to the detector's frame plays an important role. As discussed in the Introduction, the corresponding frequency modulation differs for Minkowski and Rindler modes as observed in accelerated and static frames, respectively. Furthermore, the accelerated detector perceives a local horizon while the static one does not. These classical distinctions were not reflected in the choice of the field vacuum state, as the existence of the vacuum is a purely quantum concept.  Therefore, we consider a coherent state, which provides the closest quantum analogue of a classical field configuration.

Both the excitation and de-excitation probabilities of the detector, corresponding to the aforesaid two scenarios, have been evaluated when it is interacting with a single field mode living in the coherent state. The analysis was done in $(1+1)$ and $(3+1)$-spacetime dimensions. While the analytic expressions in $(1+1)$-dimensions are valid for the full parameter space, the same in $(3+1)$-dimensions are obtained by imposing conditions $\omega/a\ll1$ and $\Omega/a\ll1$. The probabilities now consist of a purely quantum part, originating from the field's vacuum, plus a classical piece. In general, we observed two completely distinct transitions for the accelerated and static detectors even if the field and detector frequencies are taken to be same. Such a mismatch is completely due to the classical contributions to the probabilities. Therefore, the reflection of symmetry in the relative motion between the detector and the field on the transition probabilities is dependent on the choice of background field state.

For a better understanding, we analysed the results in two distinct limits. First, we considered the semiclassical limit of the field state, which corresponds to the quantum detector's response to the classical field. Here, as expected, the vacuum parts in the transition probabilities completely vanish while the classical pieces survive. So there is no sign of equivalence in this limit. We further impose the condition for the detector's close-to-classical behaviour on the transition probabilities. These results can be interpreted as the response of the classical detector to a classical field. Expressions were obtained at the resonance condition, i.e. $\omega=\Omega$. Interestingly, in $(1+1)$-dimensions, the transition probabilities in the two scenarios coincide, reflecting the equivalence between the responses of accelerated and static detectors to the Minkowski and the Rindler field modes, respectively. However, such is not the case within the region of validity of our expressions in $(3+1)$-dimensions.
   
The analysis in $(3+1)$ dimensions can be extrapolated to get an idea of the same in $(2+1)$-dimensions. In $(3+1)$ dimensions, the Minkowski and the Rindler mode functions are given by Eqs. (\ref{u3M-wmD}) and (\ref{u3R-wmD}), respectively. Replacing the transverse factor $e^{ik_\perp\cdot x_\perp}$ with $e^{ik_y y}$ (no $x$-coordinate) and changing the normalization factor appropriately, one can obtain the respective mode functions in  $(2+1)$ dimensions. Since the derivation of transition probabilities for $(2+1)$ dimensions proceeds identically as in $(3+1)$  dimensions, the expression for $P_{vac}$ remains unchanged apart from the normalization factor. Similarly, $P_{\alpha}$ has the same structure as in $(3+1)$ dimensions, with the replacement $e^{ik_\perp\cdot x_\perp}\rightarrow e^{ik_y y}$ together with the corresponding normalization constant. Hence, the outcomes in $(2+1)$ dimensions will be similar to those in $ (3+1)$ dimensions, signifying the role of the transverse coordinates.

In general, inequivalence of the detector's responses provides the signature of the natural distinguishability of the Minkowski and Rindler field coherent modes between accelerated and static frames, respectively, which is absent when the field is in its vacuum state. In the classical field limit, this distinguishability is further bolstered because the behaviour of the quantum detector in the classical field depends on the choice of frame. The similarity of the results in $(1+1)$-dimensions when both the objects are close to classical implies that the indistinguishability of the phenomenon in two scenarios is more expected in the classical regime, while quantum mechanical events can be different. In this regard, the detector's classicality is crucial. However, since we do not see any such signature in $(3+1)$ dimensions, a concrete comment on this can not be made at the present stage. Further, the EEP connects Minkowski static and accelerated observers to freely falling and static observers, respectively, in black hole spacetime. Therefore, the present results signify a similar situation for a black hole spacetime as well.

\vskip 3mm
\begin{acknowledgments}{\it Acknowledgments.}-- The research of BRM is partially supported by a START-UP RESEARCH GRANT (No. SG/PHY/P/BRM/01) from the Indian Institute of Technology Guwahati, India.
\end{acknowledgments}

\begin{widetext}
\section*{Appendices}
\begin{appendix}

\section{Transition probabilities without boundary in the $(1+1)$ dimensions}\label{Appendix_NM_1}
In this Appendix, we present the detector transition probabilities from a coherent field state in the $(1+1)$ dimensions of spacetime in the absence of a
boundary (mirror). Here we consider two scenarios: (i) an accelerating atom observing the Minkowski vacuum, and (ii) a stationary atom observing the Rindler vacuum.

\subsection{The atom is uniformly accelerating}\label{Appendix_NM_1a}
The Minkowski field modes are known as $u^{M}_{k}(x)=e^{-i\omega t+ikz}/\sqrt{4\pi\omega}$. Using the trajectories in Eq. (\ref{xtD}) it can be expressed as
\begin{eqnarray}\label{u-m-1}
    u_{k}^{M}(x)
    =\frac{1}{\sqrt{4\pi\omega}}e^{\frac{ i\omega}{a}e^{-a\tau} }\,.
\end{eqnarray}
The detector excitation and de-excitation probabilities from field vacuum can be calculated using Eq. (\ref{p-vac}) and (\ref{u-m-1}), and are already known as \cite{Kumawat:2024kul}
\begin{eqnarray}\label{P-e-nm1}
&&P_{vac}(\Omega)
  =\frac{\lambda^{2}\,\hbar_{f}}{2\omega a \Omega\,\hbar_{d}}\frac{1}{e^{\frac{2\pi\Omega }{a}}-1} \,; \\
  \label{P-d-nm1}
&&P_{vac}(-\Omega)
  =\frac{\lambda^{2}\,\hbar_{f}}{2\omega a \Omega\,\hbar_{d}}\frac{1}{1-e^{-\frac{2\pi\Omega }{a}}}  \,.
\end{eqnarray}

Now, from Eq. (\ref{p-c}) and Eq. (\ref{u-m-1}), one can calculate non-vacuum contributions to the excitation and de-excitation probabilities as
\begin{eqnarray}{\label{eq: P-NM-a-e-1a}}
   P_{\alpha}(\Omega)
   &=& \frac{\lambda^{2}\alpha_{k}^{2}\,\hbar_{f}}{4\pi\omega\,\hbar_{d}}\Bigg|\int_{-\infty}^{\infty}d\tau\; e^{-i\Omega\tau}    \biggl(e^{\frac{ i\omega}{a}e^{-a\tau} }+e^{-\frac{ i\omega}{a}e^{-a\tau} }\biggl)\Bigg|^{2}\nonumber\\
   &=&  \frac{\lambda^{2}\alpha_{k}^{2}\,\hbar_{f}}{4\pi\omega\,\hbar_{d}}\frac{1}{a^{2}}\lvert  \int^{\infty}_{0} dy\; \;y^{\frac{i\Omega }{a}-1}    \big[e^{\frac{i\omega y}{a}}+e^{-\frac{i\omega y}{a}}\big]\rvert^{2}\nonumber\\
&=& \frac{\lambda^{2}\alpha_{k}^{2}\,\hbar_{f}}{2\omega a \Omega\,\hbar_{d}}\coth{\bigg(\frac{\pi\Omega}{2a}\bigg)}\,;
\end{eqnarray}
Here, we used the coordinate transformation $e^{-a\tau}=y$ to evaluate the integrals. Then we used the integral relation in Eq. (\ref{GammaR}) to obtain the final expression.

However, the vacuum contribution to the excitation and de-excitation probabilities follows $P_{vac}(\Omega)/P_{vac}(-\Omega)=e^{-2\pi\Omega/a}$.\\

\subsection{The atom is static}\label{Appendix_NM_1s}
Here we consider the atom is static at $z=z_{0}$. In the Rindler region, the mode functions of the Rindler quantum field is known as $u_{k}^{R}(x)=\frac{e^{i\omega (\xi-\eta)}}{\sqrt{4\pi\omega}}$. Using the trajectories in Eq. (\ref{xtD}), one can express these modes in terms of Minkowski coordinates as \cite{Kumawat:2024kul} 
\begin{eqnarray}\label{u-m-1s}
  u_{k}^{R}(x)
=\frac{(a(z-t))^{\frac{i\omega}{a}}\theta(z-t)}{\sqrt{4\pi\omega}}\,.
    \end{eqnarray}
Then, the excitation and de-excitation probabilities from field vacuum can be calculated using Eq. (\ref{p-vac}) and (\ref{u-m-1s}), and are already known as \cite{Kumawat:2024kul}
\begin{eqnarray}\label{P-e-nm1s}
     &&P^{}_{vac}(\Omega)=\frac{\lambda^{2}\,\hbar_{f}}{2 a \Omega^{2}\,\hbar_{d}}\frac{1}{e^{\frac{2\pi\omega }{a}}-1} \,; \\
     \label{P-d-nm1s}
&&     P^{}_{vac}(-\Omega)  =\frac{\lambda^{2}\,\hbar_{f}}{2 a \Omega^{2}\,\hbar_{d}}\frac{1}{1-e^{-\frac{2\pi\omega }{a}}}  \,.
\end{eqnarray}

Now, from Eq. (\ref{p-c}) and Eq. (\ref{u-m-1s}), one can calculate non-vacuum contributions to the excitation and de-excitation probabilities as
\begin{equation}\label{ }\begin{aligned}
    P^{}_{\alpha} (\Omega)
     &= \frac{\lambda^{2}\alpha_{k}^{2}\,\hbar_{f}}{4\pi\omega\,\hbar_{d}}\Big|\int_{-\infty}^{\infty}dt\; e^{-i\Omega t}  \theta(z_{0}-t)  \big[(a(z_{0}-t))^{\frac{i\omega}{a}}+(a(z_{0}-t))^{-\frac{i\omega}{a}}\big]\Big|^{2}\\
    &=\frac{\lambda^{2}\alpha_{k}^{2}\,\hbar_{f}}{4\pi\omega a^{2}\,\hbar_{d}}\Big|\int_{0}^{\infty}du\, e^{i\Omega\left(\frac{u}{a}-z_{0}\right)}\Big[
   u^{\frac{i\omega}{a}}+u^{-\frac{i\omega}{a}} \Big]\Big|^{2}
    \\
  & =   \frac{\lambda^{2}\alpha_{k}^{2}\omega\,\hbar_{f}}{4\pi a^{2}\Omega^{2}\,\hbar_{d}}\;   \Big|
   e^{\frac{\pi\omega }{2a}}\; 
       \resizebox{0.12\textwidth}{!}{${\scriptsize \biggl(\frac{\Omega}{a}\biggl)^{\frac{i\omega}{a}}\;\Gamma\biggl(-\frac{i\omega }{a}\biggl) }$}
       - e^{-\frac{\pi\omega }{2a}}\; 
        \resizebox{0.12\textwidth}{!}{${\scriptsize \biggl(\frac{\Omega}{a}\biggl)^{-\frac{i\omega}{a}}\;\Gamma\biggl(\frac{i\omega }{a}\biggl)}$} \Big|^{2}\,.~
\end{aligned}\end{equation}
Here we utilized change of variable $u=a(z_{0}-t)$ to evaluate the integrals, and used the relation (\ref{GammaR}) to obtain the final expression.

These expressions can be further simplified as
\begin{eqnarray}{\label{eq: P-NM-a-e-1s}}
     && P^{}_{\alpha} (\Omega)
     =   \frac{\lambda^{2}\alpha_{k}^{2}\,\hbar_{f}}{4 a\Omega^{2}\,\hbar_{d}\sinh\frac{\pi\omega}{a}}\;   \Big|
   e^{\frac{\pi\omega }{2a}} e^{i\phi}
       - e^{-\frac{\pi\omega }{2a}}e^{-i\phi} \Big|^{2}\,
\end{eqnarray}

Here we defined $\phi=Arg\big[\big(\frac{\Omega}{a}\big)^{\frac{i\omega}{a}}\;\Gamma\big(-\frac{i\omega }{a}\big)\big]$. 
However, the vacuum contributions in excitation and de-excitation probabilities follow the relation $P_{vac}(\Omega)/P_{vac}(-\Omega)=e^{-\frac{2\pi\omega}{a}}$.


\section{ Transition probabilities with boundary in the $(3+1)$ dimensions}\label{Appendix_WM_3}

In this appendix, we provide the details of the calculations leading to the non-vacuum contributions to the excitation and de-excitation probabilities with a reflecting boundary in $(3+1)$ dimensions for the two configurations considered in the main text.

\subsection{The atom is uniformly accelerating}\label{Appendix_WM_3a}
For asymptotically large accelerations ($\frac{\omega}{a}<<1$), one can utilise the asymptotic expansion for the Bessel function \cite{Gradshteyn, Crispino:2007eb, Kumawat:2024kul}
\begin{eqnarray}\label{besselKasym}
    K_{\mu}(x)\approx \frac{1}{2}\left[\left(\frac{x}{2}\right)^{\mu}\Gamma(-\mu)+\left(\frac{x}{2}\right)^{-\mu}\Gamma(\mu)\right]\,;
\end{eqnarray}

The small-argument expansion of the modified Bessel function $K_{\mu}(x)$ in Eq.~(\ref{besselKasym}) is valid for a fixed value of the order $\mu$. For purely imaginary order $\mu=i\nu$, the expansion of the modified Bessel function $K_{i\nu}(x)$ is oscillatory in a neighbourhood of $x=0$. While well-defined for each fixed $\nu$, this expression is not uniformly valid over an unbounded range of $\nu$, due to increasingly rapid oscillations as $|\nu|\to\infty$ (see \cite{olver1997asymptotics,dunster1990bessel}). More precisely, the expansion is uniform in a bounded region of $\nu$ satisfying $|\nu\ln(x/2)|\ll 1$.
In our analysis, we have used this asymptotic expansion in the small $\omega/a$ regime. Therefore the expansion is applied in a chosen range of $\Omega$, $\omega$ and $a$ such that the condition
\begin{equation}
\left| \frac{\Omega}{a} \ln\left( \frac{\omega \sin \theta}{2a} \right) \right| \ll 1~,
\end{equation}
is satisfied, so that the expansion remains uniform in $\Omega$.

Hence, in that limit, our detector excitation and de-excitation probabilities in Eq. (\ref{P-a-e-3a-A}) can be further simplified and expressed as

\begin{eqnarray}
    &&P_{\alpha}(\Omega)
    =\frac{\lambda^2\alpha_k^2\,\hbar_{f}}{16\pi^3a^2\omega\,\hbar_{d}} \Big|
    e^{-ik_{\perp}\cdot x_{\perp}+\frac{\pi\Omega}{2a}} 
   \Big[ e^{ik_{z}z_{0}}\,\Gamma\left(-\frac{i\Omega}{a}\right)\left(\frac{\omega(1-\cos{\theta})}{2a}\right)^{\frac{i\Omega}{a}}
   +e^{ik_{z}z_{0}}\,\Gamma\left(\frac{i\Omega}{a}\right)\left(\frac{\omega(1+\cos{\theta})}{2a}\right)^{-\frac{i\Omega}{a}}\nonumber\\
   &&~~~~~~~~~~~~~~~~~~~~~~~~~~~~~~~~~
   -e^{-ik_{z}z_{0}}\,\Gamma\left(-\frac{i\Omega}{a}\right)\left(\frac{\omega(1+\cos{\theta})}{2a}\right)^{\frac{i\Omega}{a}}+e^{-ik_{z}z_{0}}\,\Gamma\left(\frac{i\Omega}{a}\right)\left(\frac{\omega(1-\cos{\theta})}{2a}\right)^{-\frac{i\Omega}{a}}\Big]\nonumber\\
   &&~~~~~~~~~~~~~~~
+e^{ik_{\perp}\cdot x_{\perp}-\frac{\pi\Omega}{2a}} \Big[ e^{-ik_{z}z_{0}}\,\Gamma\left(-\frac{i\Omega}{a}\right)\left(\frac{\omega(1-\cos{\theta})}{2a}\right)^{\frac{i\Omega}{a}}+e^{-ik_{z}z_{0}}\,\Gamma\left(\frac{i\Omega}{a}\right)\left(\frac{\omega(1+\cos{\theta})}{2a}\right)^{-\frac{i\Omega}{a}}\nonumber\\
&&~~~~~~~~~~~~~~~~~~~~~~~~~~~~~~~
-e^{ik_{z}z_{0}}\,\Gamma\left(-\frac{i\Omega}{a}\right)\left(\frac{\omega(1+\cos{\theta})}{2a}\right)^{\frac{i\Omega}{a}}+ e^{ik_{z}z_{0}}\,\Gamma\left(\frac{i\Omega}{a}\right)\left(\frac{\omega(1-\cos{\theta})}{2a}\right)^{-\frac{i\Omega}{a}}\Big]\Big|^2.
\end{eqnarray}

The expression can be further simplified by representing the phase factors in terms of the quantities $\psi_{1}$ and $\psi_{2}$, resulting in Eq.~(\ref{P-a-e-3a}).

\subsection{The atom is static}\label{Appendix_WM_3s}
For a static atom in the presence of a uniformly accelerating mirror, the non-vacuum contributions to the excitation and de-excitation probabilities are determined by Eqs.~(\ref{p-c}) and (\ref{u3Rm}) as follows:

\begin{eqnarray}
       P_{\alpha}(\Omega)&=&\frac{\lambda^2\alpha_k^2\,\hbar_{f}}{\hbar_{d}}\Big|\int_{-\infty}^{\infty}dt\,e^{-i\Omega t}\Big\{(U-V)[\theta(z_0-t)\,\big(a(z_0-t)\big)^{\frac{i\omega}{a}} -\theta(z_0+t)\,\big(a(z_0+t)\big)^{-\frac{i\omega}{a}}]\,\nonumber\\
        &&~~~~~~~~~~~~~~~~~~~~~~~~~~~~~~~~~~~~~~~~ \,+\,(U-V)^{\star}[\theta(z_0-t)\,\big(a(z_0-t)\big)^{-\frac{i\omega}{a}} -\theta(z_0+t)\,\big(a(z_0+t)\big)^{\frac{i\omega}{a}}]\Big\}\Big|^2\,\nonumber\\
        &=&\frac{\lambda^2\alpha_k^2\,\hbar_{f}}{a^{2}\,\hbar_{d}}\Big|\int_{0}^{\infty}dy\,e^{i\Omega(\frac{y}{a}-z_{0})}[(U-V)y^{\frac{i\omega}{a}}+(U-V)^{\star}y^{\frac{-i\omega}{a}}]-\int_{0}^{\infty}d\tilde{y}\,e^{-i\Omega(\frac{\tilde{y}}{a}-z_{0})}[(U-V)\tilde{y}^{\frac{-i\omega}{a}}+(U-V)^{\star}\tilde{y}^{\frac{i\omega}{a}}]\Big|^{2}\,.
\end{eqnarray}

Here, we have used the change of variables $a(z_{0}-t)=y$ and $a(z_{0}+t)=\tilde{y}$ to simplify the integrations. Then, the integrations are performed using Eq. (\ref{GammaR}) and expressed in terms of Gamma functions. The amplitude is further simplified by expressing $(U-V)$ in polar form and combining the associated phase factors into the quantities $\phi$ and $\varphi_4$. Taking the modulus squared of the amplitude yields the compact expression presented in Eq.~(\ref{P-a-e-3s}).


\section{Transition probabilities without boundary in the $(3+1)$ dimensions}\label{Appendix_NM_3}
In this Appendix, we present the detector transition probabilities from a coherent field state in the $(3+1)$ dimensions of spacetime in the absence of a
boundary (mirror). Here we consider two scenarios: (i) an accelerating atom observing the Minkowski vacuum, and (ii) a stationary atom observing the Rindler vacuum.

\subsection{The atom is uniformly accelerating}\label{Appendix_NM_3a}
In $(3+1)$ dimensions, the mode functions of a massless real scalar field in Minkowski spacetime are known as
\begin{equation}\label{MM3D}
    u_{k}^{M}(x)
    =\frac{e^{-i\omega t+ik_{z}z+i\vec{k}_{\perp}.\vec{x}_{\perp}}}{\sqrt{(2\pi)^3\,2\omega}}\,.
    \end{equation}
    Here we take $k_z=\omega\cos\theta$ and $k_{\perp}=(k_x,\,k_y)$. In this case, we consider acceleration along the $z$-direction. The trajectory of the detector is given by 
\begin{eqnarray}\label{xt3}
    t = \frac{e^{a\xi}}{a} \sinh{\left(a\,\eta\right)}~;~~z= \frac{e^{a\xi}}{a} \cosh{\left(a\,\eta\right)}~;~~    \vec{x}_{\perp} = const.
\end{eqnarray}
Setting $\xi=0$, coordinate time $\eta$ reduces to the detector's proper time $\tau$. 
Now one need to re-express the mode functions using the above coordinate transformation. Using Eq. (\ref{xt3}), one obtain
     \begin{eqnarray}
    &&   t-z\cos\theta = \frac{\bigl(e^{a\tau} - e^{-a\tau}\bigl)}{2a}-\frac{\bigl(e^{a\tau} + e^{-a\tau}\bigl)\cos\theta}{2a}\nonumber\\
    &&~~~~~~~~~~~~~~~~
=\frac{1}{\omega}[A\, e^{a\tau}-B\, e^{-a\tau}]\,;
    \end{eqnarray}
where $A$ and $B$ are defined in Eq.~(\ref{AB}). Then we can rewrite the mode in Eq. (\ref{MM3D}) as
    \begin{equation}\label{u3M}
    u_{k}^{M}(x)=\frac{e^{-i(Ae^{a\tau}-Be^{-a\tau})+i\vec{k}_{\perp}.\vec{x}_{\perp}}}{\sqrt{(2\pi)^3\,2\omega}}\,.
    \end{equation}

The detector excitation and de-excitation probabilities from the field vacuum can be evaluated using Eq. (\ref{p-vac}) and (\ref{u3M}), and are known as \cite{Kumawat:2024kul}
\begin{eqnarray}\label{P-e-nm3}
        &&{P}_{vac}(\Omega)
        = \frac{\lambda^2\,\hbar_{f}e^{-\frac{\pi\Omega}{a}}}{4\pi^3a^2\omega\,\hbar_{d}}\,{K}^2_{\frac{i\Omega}{a}}\Big(\frac{\omega}{a}\sin{\theta}\Big)~;\\
        \label{P-d-nm3}
     &&  {P}_{vac}(-\Omega) = \frac{\lambda^2\,\hbar_{f}e^{\frac{\pi\Omega}{a}}}{4\pi^3a^2\omega\,\hbar_{d}}\, {K}^2_{\frac{i\Omega}{a}}\Big(\frac{\omega}{a}\sin{\theta}\Big)~.
\end{eqnarray}
For the condition $\frac{\omega}{a}<<1$, one can use the asymptotic expansion for the Bessel function (\ref{besselKasym}). Then we obtain 
\begin{eqnarray}\label{besselKasymMod2}
   && \Big|{K}_{\frac{i\Omega}{a}}\bigg(\frac{\omega }{a}\sin{\theta}\bigg) \Big|^{2}=\frac{\pi}{\frac{\Omega }{a} \sinh{(\frac{\pi\Omega }{a})}}\;\cos^{2}{\varphi_3}\,.
\end{eqnarray}
Here we denoted $\varphi_3=Arg\Big[\Big(\frac{\omega\sin{\theta}}{2a}\Big)^{\frac{i \Omega }{a}}\Gamma\Big(-\frac{i\Omega }{a}\Big)\Big]$. Hence, in that limit, our detector excitation and de-excitation probabilities from the field vacuum can be further simplified as \cite{Kumawat:2024kul}
\begin{eqnarray}\label{P-e-nm3-2}
 &&{P}_{vac}(\Omega)
        = \frac{\lambda^2\,\hbar_{f}}{2\pi^2\,a\,\omega\,\Omega\,\hbar_{d}}\frac{\cos^2\varphi_3}{e^{\frac{2\pi\Omega}{a}}-1}~;\\
        \label{P-d-nm3-2}
 &&{P}_{vac}(-\Omega)
        = \frac{\lambda^2\,\hbar_{f}}{2\pi^2\,a\,\omega\,\Omega\,\hbar_{d}}\frac{\cos^2\varphi_3}{1-e^{-\frac{2\pi\Omega}{a}}}~.
\end{eqnarray}
Now, from Eq. (\ref{p-c}) and (\ref{u3M}), one can calculate non-vacuum contributions in the excitation and de-excitation probabilities as
\begin{eqnarray}
    P^{}_{\alpha}  (\Omega) 
   &=& \frac{\lambda^2\alpha_{k}^{2}\,\hbar_{f}}{(2\pi)^{3}2\omega\,\hbar_{d}}
   \Big|e^{i\vec{k}_{\perp}.\vec{x}_{\perp}}\int_{-\infty}^{\infty}d\tau\;e^{-i\Omega\tau}\;e^{-i(Ae^{a\tau}-Be^{-a\tau})}+e^{-i\vec{k}_{\perp}.\vec{x}_{\perp}}\int_{-\infty}^{\infty}d\tau\; e^{-i\Omega\tau}\;e^{i(Ae^{a\tau}-Be^{-a\tau})} \Big|^{2}\,\nonumber\\
   &=&\frac{\lambda^2\alpha_{k}^{2}\,\hbar_{f}}{(2\pi)^{3}2\omega a^{2}\,\hbar_{d}}
   \Big|e^{i\vec{k}_{\perp}.\vec{x}_{\perp}}\int_{0}^{\infty}dy\; y^{-\frac{i\Omega}{a}-1}\;e^{-(e^{\frac{i\pi}{2}}Ay+e^{\frac{-i\pi}{2}}\frac{B}{y})} + e^{-i\vec{k}_{\perp}.\vec{x}_{\perp}}\int_{0}^{\infty}dy\; y^{-\frac{i\Omega}{a}-1}\;e^{-(e^{\frac{-i\pi}{2}}Ay+e^{\frac{i\pi}{2}}\frac{B}{y})} \Big|^{2}.
\end{eqnarray}
Here, we have used the change of variables $e^{a\tau}=y$ to simplify the expression. Then we use the following integral expression in Eq. (\ref{besselK}) to evaluate the integrals. We obtain
\begin{eqnarray}
    P^{}_{\alpha}  (\Omega) 
   =\frac{\lambda^2\alpha_{k}^{2}\,\hbar_{f}\,  K^{2}_{\frac{i \Omega }{a}}\left(\frac{\omega\sin\theta}{a}\right)}{4\pi^{3}\omega a^{2}\,\hbar_{d}}
    \Big|e^{i\vec{k}_{\perp}.\vec{x}_{\perp}}e^{-\frac{\pi  \Omega }{2 a}} +e^{-i\vec{k}_{\perp}.\vec{x}_{\perp}}e^{\frac{\pi  \Omega }{2 a}}\Big|^{2}\,.~~~~
\end{eqnarray}

The vacuum contribution to the excitation and de-excitation probabilities follows the relation $P_{vac}(\Omega)/P_{vac}(-\Omega)=e^{-2\pi\Omega/a}$. For the condition $(\omega/a)<<1$ (using Eq. (\ref{besselKasymMod2})), one can rewrite $P_\alpha(\Omega)$ as
\begin{eqnarray}\label{eq: P-NM-a-e-3a}
    P^{}_{\alpha}(\Omega)   
   =\frac{\lambda^2\alpha_{k}^{2}\,\hbar_{f} \cos^2\varphi_3}{4\pi^{2}\omega a\Omega\,\hbar_{d}\sinh{\frac{\pi\Omega}{a}}}
    \Big|e^{i\vec{k}_{\perp}.\vec{x}_{\perp}}e^{\frac{\pi  \Omega }{2 a}} +e^{-i\vec{k}_{\perp}.\vec{x}_{\perp}}e^{-\frac{\pi  \Omega }{2 a}}\Big|^{2}.~~~~
\end{eqnarray}

\subsection{A static atom}\label{Appendix_NM_3s}
In (3 + 1) dimensions, the mode functions of massless real
scalar fields in Rindler spacetime are known as \cite{Crispino:2007eb}
\begin{equation}
    u^{R}_{k}(x) =\sqrt{\frac{\sinh{\frac{\pi\omega}{a}}}{4\pi^4a}}\,{K}_{\frac{i\omega}{a}}\bigg(\frac{k_{\perp}e^{a\,\xi}}{a}\bigg)~e^{i\,\vec{k}_{\perp}\cdot\vec{x}_{\perp}-i\,\omega\,\eta}~. 
\end{equation}
In the $k_{\perp}e^{a\xi}/a<<1$ limit using Eq. (\ref{besselK}), one can rewrite the mode function as 
\begin{eqnarray}\label{u3R}
        u^{R}_{k} &=&\sqrt{\frac{\sinh{\frac{\pi\omega}{ a}}}{4\pi^4a}} e^{i\,\vec{k}_{\perp}\cdot\vec{x}_{\perp}}\,\biggl[   \frac{1}{2}\Gamma\biggl(-\frac{i\omega}{a}\biggl)\,\bigg(\frac{k_{\perp}}{2a}\bigg)^{\frac{i\omega}{a}} e^{-i\omega(\eta-\xi)}\,+\,\frac{1}{2}\Gamma\biggl(\frac{i\omega}{a}\biggl)\,\bigg(\frac{k_{\perp}}{2a}\bigg)^{-\frac{i\omega}{a}} e^{-i\omega(\eta+\xi)}\biggl]~\nonumber\\
 &=&U\,e^{-i\omega(\eta-\xi)}+V\,e^{-i\omega(\eta+\xi)}\,,
 \end{eqnarray}
 where $U$ and $V$ are defined in Eq.~(\ref{UV}).
Using the coordinate transformation in Eq.  (\ref{xt3}), one can simplify the mode function in Eq. (\ref{u3R}) as
\begin{eqnarray}\label{u3R-2}
u^{R}_{k}=U\,\theta(z_0-t)\,\big(a(z_0-t)\big)^{\frac{i\omega}{a}} + V\,\theta(z_0+t)\,\big(a(z_0+t)\big)^{-\frac{i\omega}{a}}.~~
\end{eqnarray}
The detector excitation and de-excitation probabilities from the field vacuum can be evaluated using Eq. (\ref{p-vac}) and (\ref{u3R-2}), and are known as \cite{Kumawat:2024kul}

\begin{eqnarray}\label{P-e-nm3s}
        &&{P}_{vac}(\Omega)
        = \frac{\lambda^2\,\hbar_{f}}{2\pi^2\,a\,\Omega^2\,\hbar_{d}}\,\frac{\cos^2(\Omega z_0+\varphi_{ex})}{e^{\frac{2\pi\omega}{a}}-1}~;\\
        \label{P-d-nm3s}
     &&  {P}_{vac}(-\Omega) = \frac{\lambda^2\,\hbar_{f}}{2\pi^2\,a\,\Omega^2\,\hbar_{d}}\,\frac{\cos^2(\Omega z_0-\varphi_{ex})}{1-e^{-\frac{2\pi\omega}{a}}}~.
\end{eqnarray}
Here we defined $\varphi_{ex}=Arg[(\frac{k_{\perp}}{2\Omega})^{-\frac{i\omega}{a}}]$. 

Now, from Eq. (\ref{p-c}) and (\ref{u3R-2}), one can calculate non-vacuum contributions to the excitation and de-excitation probabilities as
\begin{equation}
    \begin{aligned}
         P_{\alpha}(\Omega)&=\frac{\lambda^2\alpha_{k}^{2}\,\hbar_{f}}{\,\hbar_{d}}\big|\int_{-\infty}^{z_{0}}dt e^{-i\Omega t} \big(U\big(a(z_{0}-t)\big)^{\frac{i\omega }{a}}+U^{\star}\big(a(z_{0}-t)\big)^{-\frac{i\omega }{a}}\big) +\int_{-z_{0}}^{\infty}dt e^{-i\Omega t} \big(V\big(a(z_{0}+t)\big)^{-\frac{i\omega }{a}}+V^{\star}\big(a(z_{0}+t)\big)^{\frac{i\omega }{a}}\big)\big|^{2}\\
       &=\frac{\lambda^2\alpha_{k}^{2}\,\hbar_{f}}{a^{2}\,\hbar_{d}}\big|\int_{0}^{\infty}dy e^{i\Omega (\frac{y}{a}-z_{0})} \big(Uy^{\frac{i\omega }{a}}+U^{\star}y^{-\frac{i\omega }{a}}\big)\,+\,\int_{0}^{\infty}d\tilde{y} e^{-i\Omega (\frac{\tilde{y}}{a}-z_{0})} \big(V\tilde{y}^{-\frac{i\omega }{a}}+V^{\star}\tilde{y}^{\frac{i\omega }{a}}\big)\big|^{2}
    \end{aligned}
\end{equation}
Here, we have used the change of variables $a(z_{0}-t)=y$ and $a(z_{0}+t)=\tilde{y}$ to simplify the expressions. Then the integrations are performed using Eq. (\ref{GammaR}). Further simplifying, we obtained the following:
\begin{equation}\label{eq: P-NM-a-e-3s}
    \begin{aligned}
         P_{\alpha}(\Omega)&=\frac{\lambda^2\alpha_{k}^{2}\,\hbar_{f}}{4\pi^2a\Omega^2\,\hbar_{d}\sinh{(\frac{\pi\omega}{a})}}
\big|-e^{-\frac{\pi\omega }{2a}+i\,\vec{k}_{\perp}\cdot\vec{x}_{\perp}}\cos{(\Omega z_0+\varphi_{ex})}+e^{\frac{\pi\omega }{2a}-i\,\vec{k}_{\perp}\cdot\vec{x}_{\perp}}\cos{(\Omega z_0-\varphi_{ex})}\big|^{2}\,.
    \end{aligned}
\end{equation}

Also note that the vacuum contributions in excitation and de-excitation probabilities follow the relation $P_{vac}(\Omega)/P_{vac}(-\Omega)=e^{-\frac{2\pi\omega}{a}}$ only if $z_0=0$.


\section{Semiclassical limits of the field and the detector in $(1+1)$-dimensions in the absence of boundary}\label{AppB1}

In this appendix, we first analyze the effect of the semiclassical limit of the field on the transition probabilities of both an accelerated atom and a static atom. Subsequently, we take the semiclassical limit of the detector on top of this field limit.

First, in the classical field limit,  we take $\hbar_{f}\to 0$ while keeping $\alpha_{k}^{2}\hbar_{f}$ finite. If $f(k)$ is the amplitude of the classical wave mode, it is related to the coherent-state amplitude $\alpha_k$ through $\alpha_k^{2}\hbar = \beta\,\omega\,f^{2}$. In this limit, the vacuum contribution $P_{\text{vac}}(\pm\Omega)$ vanishes, whereas the effective-field contribution $P_{\alpha}(\pm\Omega)$ remains nonzero for both accelerating and static atoms.
Finally, we impose the semiclassical limit on the detector. 
Taking the detector semiclassical limit $\hbar_d\to0$, the effective-field contribution to the transition probability of the accelerated atom, given in Eq.~(\ref{eq: P-NM-a-e-1a}), reduces to
\begin{equation}{\label{eq:P-nm-a-e-1a-Climit-F&D}}
P_{\alpha}
=\frac{\lambda^{2}\beta f^{2}}{2a\,\Delta E}.
\end{equation}
The same result is obtained for the static atom from Eq.~(\ref{eq: P-NM-a-e-1s}). In this limit, the phase $\phi$ vanishes since $\Gamma(\pm i\infty)=0$.


\section{Semiclassical Limits of transition probabilities under the resonance condition in $(3+1)$-dimensions}

In this appendix, we first take the semiclassical limit of the field and subsequently impose the resonance condition to simplify the transition probabilities of the atom, both in the presence and absence of a reflecting boundary, in the large-acceleration limit.

\subsection{With Mirror}\label{Appendix: WM_3}

\subsubsection{The atom is uniformly accelerating}\label{sec:P-a-e-3a}
In the large-acceleration limit ($\omega/a \ll 1$ and $\Omega/a \ll 1$), the quantities $\kappa_{1}$ and
$\kappa_{2}$ can be approximated as
$\kappa_{1}\simeq -i\,a/\Omega-\ln[(\omega+k_{z})/2a]-\gamma$ and
$\kappa_{2}\simeq i\,a/\Omega-\ln[(\omega-k_{z})/2a]-\gamma$, where
$\gamma$ is the Euler--Mascheroni constant.  As a result, the associated
phases reduce to
$\psi_{1}\simeq -\pi/2-(\Omega/a)[\ln((\omega+k_{z})/2a)+\gamma]$ and
$\psi_{2}\simeq \pi/2+(\Omega/a)[\ln((\omega-k_{z})/2a)+\gamma]$.  Within
this approximation, Eq.~(\ref{P-a-e-3a}) can be rewritten as
\begin{equation}
    \begin{aligned}
     &  P_{\alpha}(\Omega)  =\frac{\lambda^2\alpha_k^2\hbar_f}{4\pi^2a\omega\Omega\,\hbar_{d}\,\sinh\frac{\pi\Omega}{a}}  \Big|
    e^{-ik_{\perp}\cdot x_{\perp}+\frac{\pi\Omega}{2a}} 
   \Big( \cos{(k_{z}z_{0}+\frac{\Omega}{a}[\ln{\left(\frac{\omega-k_{z}}{2a}\right)}+\gamma])}-
   \cos{(k_{z}z_{0}-\frac{\Omega}{a}[\ln{\left(\frac{\omega+k_{z}}{2a}\right)}+\gamma])} \Big)\\&
  ~~~~~~~~~~~~~~~~~~~~~~~~~~~~~~~~~~~~~~~~~~~~~~~~~~ 
    -e^{ik_{\perp}\cdot x_{\perp}-\frac{\pi\Omega}{2a}}\Big(\cos{(k_{z}z_{0}+\frac{\Omega}{a}[\ln{\left(\frac{\omega+k_{z}}{2a}\right)}+\gamma])}-\cos{(k_{z}z_{0}-\frac{\Omega}{a}[\ln{\left(\frac{\omega-k_{z}}{2a}\right)}+\gamma])}\Big)\Big|^2\,.
    \end{aligned}
\end{equation}
The classical-field regime is obtained by taking the limit $\hbar_f\to0$ while keeping $\alpha_k^{2}\hbar_f$ fixed, which relates the coherent-state amplitude $\alpha_k$ to the classical mode amplitude $f(k)$ through $\alpha_k^{2}\hbar=\beta\,\omega f^{2}$.  Imposing the resonance condition $\Omega=\omega$ and writing the detector frequency as $\Omega=\Delta E/\hbar_d$, the large-acceleration regime is characterized by $\Delta E/(\hbar_d a)\ll1$.  In this limit, the effective-field contribution $P_{\alpha}$ simplifies as
\begin{eqnarray}
P_{\alpha}(\Omega) &\simeq& \frac{\lambda^2\beta \,f^{2}}{4\pi^2a\Delta E}
        \Bigg|e^{-ik_{\perp}\cdot x_{\perp}} \frac{e^{\frac{\pi \Delta E}{2 \hbar_d a}}}{\sinh^{1/2}\!\left(\frac{\pi \Delta E}{\hbar_d a}\right)}\Bigg[ \cos\Big\{\frac{\Delta E}{\hbar_d} z_0\cos\theta +\frac{\Delta E}{\hbar_d a}\Big[\ln{\left(\frac{\Delta E}{2\hbar_d a}(1-\cos\theta)\right)}+\gamma\Big]\Big\} \nonumber\\
        &&~~~~~~~~~~~~~~~~~~~~~~~ ~~~~~~~~~~~~~~~~~~~~~~~~~~~~~~~-\cos\Big\{\frac{\Delta E}{\hbar_d} z_0\cos\theta -\frac{\Delta E}{\hbar_d a}\Big[\ln{\left(\frac{\Delta E}{2\hbar_d a}(1+\cos\theta)\right)}+\gamma\Big]\Big\} \Bigg] \nonumber\\
        &&~~~~~~~~~~~~~~~~~~-e^{ik_{\perp}\cdot x_{\perp}}\frac{e^{-\frac{\pi \Delta E}{2 \hbar_d a}}}{\sinh^{1/2}\!\left(\frac{\pi \Delta E}{\hbar_d a}\right)}\Bigg[ \cos\Big\{\frac{\Delta E}{\hbar_d} z_0\cos\theta +\frac{\Delta E}{\hbar_d a}\Big[\ln{\left(\frac{\Delta E}{2\hbar_d a}(1+\cos\theta)\right)}+\gamma\Big]\Big\} \nonumber\\
        &&~~~~~~~~~~~~~~~~~~~~~~~~~~~~~~~~~~~~~~~~~~~~~~~~~~~~~~~~-\cos\Big\{\frac{\Delta E}{\hbar_d} z_0\cos\theta-\frac{\Delta E}{\hbar_d a}\Big[\ln{\left(\frac{\Delta E}{2\hbar_d a}(1-\cos\theta)\right)}+\gamma\Big]\Big\} \Bigg]
        \Bigg|^2 \,\nonumber\\
        &\simeq& \frac{\lambda^2\beta \,f^{2}}{4\pi^2a\Delta E}\Bigg|e^{-ik_{\perp}\cdot x_{\perp}}\left[\sqrt{\frac{\hbar_d a}{\pi \Delta E}}+ \frac{1}{2}\sqrt{\frac{\pi \Delta E}{\hbar_d a}}\right] \times\Bigg\{-\sin\!\left(\frac{\Delta E}{\hbar_d} z_0\cos\theta\right)\frac{\Delta E}{\hbar_d a}\left[2\gamma+\ln\!\left(\frac{\Delta E^{2}}{4\hbar_d^{2}a^{2}}\sin^{2}\theta\right)\right]\Bigg\} \nonumber\\
        &&~~~~~~~~~~~~~~~~~~~~~-e^{ik_{\perp}\cdot x_{\perp}}\left[\sqrt{\frac{\hbar_d a}{\pi \Delta E}}- \frac{1}{2}\sqrt{\frac{\pi \Delta E}{\hbar_d a}}\right]  \Bigg\{-\sin\!\left(\frac{\Delta E}{\hbar_d} z_0\cos\theta\right)\frac{\Delta E}{\hbar_d a}\left[2\gamma+\ln\!\left(\frac{\Delta E^{2}}{4\hbar_d^{2}a^{2}}\sin^{2}\theta\right)\right]\Bigg\}\Bigg|^2 \,\nonumber\\
        &\simeq& \frac{\lambda^2\beta \,f^{2}}{4\pi^2a\Delta E}\Bigg|\sin\!\left(\frac{\Delta E}{\hbar_d} z_0\cos\theta\right)\left(\frac{2\Delta E}{\hbar_d a}\right)\left[\gamma+\ln\!\left(\frac{\Delta E}{2\hbar_d a}\sin\theta\right)\right]  \left[\sqrt{\frac{\hbar_d a}{\pi \Delta E}}\,2i\sin(\vec{k}_{\perp}\!\cdot\!\vec{x}_{\perp})-\frac{1}{2}\sqrt{\frac{\pi \Delta E}{\hbar_d a}}\,2\cos(\vec{k}_{\perp}\!\cdot\!\vec{x}_{\perp})\right]\Bigg|^2    
        \,~.\nonumber\\
\end{eqnarray}   
In obtaining the above result, we have used $k_{z}=\omega\cos\theta$ in the first step.  We have also employed the large-acceleration expansions $e^{\pm\frac{\pi \Delta E}{2 \hbar_{d} a}}/{\sinh^{1/2}\!\left(\frac{\pi \Delta E}{\hbar_{d} a}\right)}\simeq\sqrt{{\hbar_{d} a}/{\pi \Delta E}}\pm \frac{1}{2}\sqrt{{\pi \Delta E}/{\hbar_{d} a}}$ and $\cos\!\left(\frac{\Delta E}{\hbar} z_0 \cos\theta\pm\frac{\Delta E}{\hbar a}\left[\ln\!\left(\frac{\Delta E}{2\hbar a}(1\pm\cos\theta)\right)+\gamma\right]\right)\simeq\cos\!\left(\frac{\Delta E}{\hbar} z_0\cos\theta\right)\mp\sin\!\left(\frac{\Delta E}{\hbar} z_0\cos\theta\right)\frac{\Delta E}{\hbar a}\left[\ln\!\left(\frac{\Delta E}{2\hbar a}(1\pm\cos\theta)\right)+\gamma\right]$,  which are valid for $\Delta E/(\hbar_d a)\ll1$.  Furthermore, in the third step, we have neglected the subleading term inside the modulus in comparison with the leading term in the limit $\Delta E/(\hbar_d a)\ll1$, which directly leads to the final simplified expression in Eq.~(\ref{P-a-e-3a-CL_F&D}).

\subsubsection{The atom is static }\label{sec:P-a-e-3s}

In the large-acceleration limit ($\omega/a \ll 1$ and $\Omega/a \ll 1$), the phases $\phi$ and $\varphi_{4}$ can be approximated as $\phi\simeq \mathrm{Arg}[\,i\,a/\omega-\ln(\Omega/a)-\gamma] \simeq \pi/2+(\omega/a)[\ln(\Omega/a)+\gamma]$ and $\varphi_{4}\simeq \mathrm{Arg}[\,i\,a/\omega-\ln(k_{\perp}/2a)-\gamma] \simeq \pi/2+(\omega/a)[\ln(k_{\perp}/2a)+\gamma]$, where $\gamma$ denotes the Euler-Mascheroni constant.  Within this approximation, Eq.~(\ref{P-a-e-3s}) can be rewritten as

\begin{eqnarray}
  P_{\alpha}(\Omega)
&=&\frac{\lambda^2\alpha_k^2\hbar_f\cos^2\left(\,\frac{\omega}{a}\left[\ln{\left(\frac{k_{\perp}}{2a}\right)}+\gamma\right]\right)}{\pi^2a\Omega^2\,\hbar_{d}\sinh\frac{\pi\omega}{a}}
\Big|e^{-\frac{\pi\omega}{a}+ik_{\perp}\cdot x_{\perp}}\cos{\left(\Omega z_0+\,\frac{\omega}{a}\left[\ln{\left(\frac{\Omega}{a}\right)}+\gamma\right]\right)}\,\nonumber\\
&&~~~~~~~~~~~~~~~~~~~~~~~~~~~~~~~~~~~~~~~~~~~~~~~~~~~~~~\,-\,e^{\frac{\pi\omega}{a}-ik_{\perp}\cdot x_{\perp}}\cos{\left(\Omega z_0-\,\frac{\omega}{a}\left[\ln{\left(\frac{\Omega}{a}\right)}+\gamma\right]\right)}\Big|^2\,.   
\end{eqnarray}
The classical-field regime is obtained by taking $\hbar_f\to0$ while keeping $\alpha_k^{2}\hbar_f$ fixed, thereby relating the coherent-state amplitude $\alpha_k$ to the classical mode amplitude $f(k)$ through $\alpha_k^{2}\hbar=\beta\,\omega f^{2}$.  With the resonance condition $\Omega=\omega$ and the detector frequency written as $\Omega=\Delta E/\hbar_d$, the large-acceleration limit is defined by $\Delta E/(\hbar_d a)\ll1$.  In this regime, the effective-field contribution $P_{\alpha}$ reduces to
\begin{eqnarray}
     P_{\alpha}(\Omega)
         &\simeq&\frac{\lambda^2\beta \,f^{2}\cos^2\left(\,\frac{\Delta E}{\hbar_{d} a}\Big[\ln{\left(\frac{\Delta E\,\sin{\theta}}{2\hbar_{d} a}\right)}+\gamma\Big]\right)}{\pi^2a\Delta E}
        \Bigg|\frac{e^{-\frac{\pi \Delta E}{2 \hbar_{d} a}}}{\sinh^{1/2}\!\left(\frac{\pi \Delta E}{\hbar_{d} a}\right)}\,e^{i\vec{k}_{\perp}\cdot \vec{x}_{\perp}}\cos{\left( \frac{\Delta E}{\hbar_{d} } z_0+\,\frac{\Delta E}{\hbar_{d} a}\Big[\ln{\left(\frac{\Delta E}{\hbar_{d} a}\right)}+\gamma\Big]\right)}\,\nonumber\\
        &&~~~~~~~~~~~~~~~~~~~~~~~~~~~~~~~~~~~~~~~~~~~~~~\,-\frac{e^{\frac{\pi \Delta E}{2 \hbar_{d} a}}}{\sinh^{1/2}\!\left(\frac{\pi \Delta E}{\hbar_{d} a}\right)}\,e^{-i\vec{k}_{\perp}\cdot \vec{x}_{\perp}}\cos{\left(\frac{\Delta E}{\hbar_{d}} z_0-\,\frac{\Delta E}{\hbar_{d} a}\Big[\ln{\left(\frac{\Delta E}{\hbar_{d} a}\right)}+\gamma\Big]\right)}\Bigg|^2\,\nonumber\\
         &\simeq&\frac{\lambda^2\beta \,f^{2}}{\pi^2a\Delta E}
        \Bigg|e^{i\vec{k}_{\perp}\cdot \vec{x}_{\perp}}\left[\sqrt{\frac{\hbar_d a}{\pi \Delta E}}- \frac{1}{2}\sqrt{\frac{\pi \Delta E}{\hbar_{d} a}}\right]\left[\cos(\frac{\Delta E}{\hbar_{d} } z_0)-\frac{\Delta E}{\hbar_{d} a}\left[\ln\!\left(\frac{\Delta E}{\hbar_{d} a}\right)+\gamma\right]\sin(\frac{\Delta E}{\hbar_{d} } z_0)\right]\,\nonumber\\
        &&~~~~~~~~~~~~~~~~~~~\,-e^{-i\vec{k}_{\perp}\cdot \vec{x}_{\perp}}\left[\sqrt{\frac{\hbar_d a}{\pi \Delta E}}+ \frac{1}{2}\sqrt{\frac{\pi \Delta E}{\hbar_d a}}\right]\left[\cos(\frac{\Delta E}{\hbar_{d} } z_0)+\frac{\Delta E}{\hbar_{d} a}\left[\ln\!\left(\frac{\Delta E}{\hbar_{d} a}\right)+\gamma\right]\sin(\frac{\Delta E}{\hbar_{d} } z_0)\right]\Bigg|^2\,\nonumber\\
         &\simeq&\frac{\lambda^2\beta \,f^{2}}{\pi^2a\Delta E} \Bigg|\cos\!\left(\frac{\Delta E}{\hbar_{d}} z_0\right)\left[\sqrt{\frac{\hbar_{d} a}{\pi \Delta E}}\,2{i\sin(\vec{k}_{\perp}\!\cdot\!\vec{x}_{\perp})}-\frac{1}{2}\sqrt{\frac{\pi \Delta E}{\hbar_{d} a}}\,2{\cos(\vec{k}_{\perp}\!\cdot\!\vec{x}_{\perp})}\right]\,\nonumber\\
        &&~~~~~~~~~~~~~-\frac{\Delta E}{\hbar_{d} a}\left(\ln\!\frac{\Delta E}{\hbar_{d} a}+\gamma\right)\sin\!\left(\frac{\Delta E}{\hbar_{d}} z_0\right)\left[\sqrt{\frac{\hbar_{d} a}{\pi \Delta E}}\,2{\cos(\vec{k}_{\perp}\!\cdot\!\vec{x}_{\perp})}-\frac{1}{2}\sqrt{\frac{\pi \Delta E}{\hbar_{d} a}}\,2{i\sin(\vec{k}_{\perp}\!\cdot\!\vec{x}_{\perp})}\right]\Bigg|^{2}\,\nonumber\\
        &\simeq&\frac{\lambda^2\beta \,f^{2}}{\pi^2a\Delta E}\Bigg|\cos\!\left(\frac{\Delta E}{\hbar_{d}} z_0\right)\left[\sqrt{\frac{\hbar_{d} a}{\pi \Delta E}}\,2{i\sin(\vec{k}_{\perp}\!\cdot\!\vec{x}_{\perp})}\right]-\left(\ln\!\frac{\Delta E}{\hbar_{d} a}+\gamma\right)\sin\!\left(\frac{\Delta E}{\hbar_{d}} z_0\right)\left[\sqrt{\frac{\Delta E}{\pi\hbar_{d} a}}\,2{\cos(\vec{k}_{\perp}\!\cdot\!\vec{x}_{\perp})}\right]\Bigg|^{2}~.              
\end{eqnarray}

In obtaining the above result, we have used $k_{\perp}=\omega\sin\theta$ in the first step.  We have also employed the large-acceleration expansions $e^{\pm\frac{\pi \Delta E}{2 \hbar_{d} a}}/{\sinh^{1/2}\!\left(\frac{\pi \Delta E}{\hbar_{d} a}\right)}\simeq\sqrt{{\hbar_{d} a}/{\pi \Delta E}}\pm \frac{1}{2}\sqrt{{\pi \Delta E}/{\hbar_{d} a}}$,$\cos\left(\,\frac{\Delta E}{\hbar_{d} a}\Big[\ln{\left(\frac{\Delta E\,\sin{\theta}}{2\hbar_{d} a}\right)}+\gamma\Big]\right)\simeq\,1$, and $\cos\!\left(\frac{\Delta E}{\hbar_{d}} z_0 \pm\frac{\Delta E}{\hbar_{d} a}\left[\ln\!\left(\frac{\Delta E}{\hbar_{d} a}\right)+\gamma\right]\right)\simeq\cos\!\left(\frac{\Delta E}{\hbar_{d}} z_0\right)\mp\sin\!\left(\frac{\Delta E}{\hbar_{d}} z_0\right)\frac{\Delta E}{\hbar_{d} a}\left[\ln\!\left(\frac{\Delta E}{\hbar_{d} a}\right)+\gamma\right]$,  all of which are valid in the regime $\Delta E/(\hbar_{d} a)\ll1$. Moreover, in this limit, the subleading terms inside the first and second square brackets in the third step are negligible compared with the dominant contributions.  Similarly, in the fourth step, the second term inside the modulus can be safely ignored relative to the leading term, which yields Eq.~(\ref{P-a-e-3s-CL_F&D}).


\subsection{Without Mirror}\label{Appendix: NM_3_Climit}

\subsubsection{The atom is uniformly accelerating}

For large acceleration we have $\omega/a\ll1$ and $\Omega/a\ll1$ and we further expand $\varphi_{3}\thickapprox\, \frac{\pi}{2}+\frac{\Omega}{a}[\ln{\frac{\omega\sin{\theta}}{2a}}+\gamma]$. Here $\gamma$ is the Euler-Mascheroni constant, which has the value $\gamma\thickapprox0.5772$.  Within this approximation, Eq.~(\ref{eq: P-NM-a-e-3a}) can be rewritten as
\begin{eqnarray}
      P^{}_{\alpha}(\Omega)   
   &=&\frac{\lambda^2\alpha_{k}^{2}\,\hbar_{f}\, \sin^2\left(\,\frac{\Omega}{a}[\ln{\left(\frac{\omega\, \sin{\theta}}{2a}\right)}+\gamma]\right)}{4\pi^{2}\omega a\Omega\,\hbar_{d}\sinh{\frac{\pi\Omega}{a}}}\,\Big|e^{i\vec{k}_{\perp}.\vec{x}_{\perp}}e^{\frac{\pi  \Omega }{2 a}} +e^{-i\vec{k}_{\perp}.\vec{x}_{\perp}}e^{-\frac{\pi  \Omega }{2 a}}\Big|^{2}.~~~~
\end{eqnarray}
The classical-field regime is obtained by taking the limit $\hbar_f\to0$ while keeping the product $\alpha_k^{2}\hbar_f$ fixed, so that the coherent-state amplitude $\alpha_k$ is related to the classical mode amplitude $f(k)$ through
$\alpha_k^{2}\hbar=\beta\,\omega f^{2}$. We impose the resonance condition $\Omega=\omega$ and express the detector frequency as $\Omega=\Delta E/\hbar_d$.  The large-acceleration regime is then characterized by $\Delta E/(\hbar_d a)\ll1$.  In this limit, the $P_{\alpha}$ can be simplified as follows:
\begin{eqnarray}
      P^{}_{\alpha}(\Omega)    
    &\simeq&\frac{\lambda^2\beta\,f^{2}\, \sin^2\left(\,\frac{\Delta E\,}{\hbar_{d}a}[\ln{\left(\frac{\Delta E\,\, \sin{\theta}}{2\hbar_{d}a}\right)}+\gamma]\right)}{4\pi^{2} a\Delta E\,}
    \Big|e^{i\vec{k}_{\perp}.\vec{x}_{\perp}}\frac{e^{\frac{\pi  \Delta E\, }{2\hbar_{d} a}}}{\sinh^{1/2}{\frac{\pi\Delta E\,}{\hbar_{d}a}}} +e^{-i\vec{k}_{\perp}.\vec{x}_{\perp}}\frac{e^{-\frac{\pi  \Delta E\, }{2\hbar_{d} a}}}{\sinh^{1/2}{\frac{\pi\Delta E\,}{\hbar_{d}a}}}\Big|^{2}\nonumber\\
    &\simeq&\frac{\lambda^2\beta\,f^{2}\, }{4\pi^{2} a\Delta E\,}
    \Big|e^{i\vec{k}_{\perp}.\vec{x}_{\perp}}\left[\sqrt{\frac{\hbar_{d} a}{\pi \Delta E}}+ \frac{1}{2}\sqrt{\frac{\pi \Delta E}{\hbar_{d} a}}\right]\left(\,\frac{\Delta E\,}{\hbar_{d}a}\left[\ln{\left(\frac{\Delta E\,\, \sin{\theta}}{2\hbar_{d}a}\right)}+\gamma\right]\right)\nonumber \\
    &&~~~~~~~~~~~~~~~~~~+e^{-i\vec{k}_{\perp}.\vec{x}_{\perp}}\left[\sqrt{\frac{\hbar_{d} a}{\pi \Delta E}}- \frac{1}{2}\sqrt{\frac{\pi \Delta E}{\hbar_{d} a}}\right]\left(\,\frac{\Delta E\,}{\hbar_{d}a}\left[\ln{\left(\frac{\Delta E\,\, \sin{\theta}}{2\hbar_{d}a}\right)}+\gamma\right]\right)\Big|^{2}\nonumber\\
    &\simeq&\frac{\lambda^{2}\beta f^{2}}{4\pi^{2} a \Delta E}\biggl|\left[\ln\!\left(\frac{\Delta E \sin\theta}{2\hbar_{d} a}\right)+\gamma\right]\left[\sqrt{\frac{\Delta E}{\pi \hbar_{d} a}}\, 2\cos(\vec{k}_{\perp}\!\cdot\!\vec{x}_{\perp})+\frac{\sqrt{\pi}}{2}\left(\frac{\Delta E}{\hbar_{d} a}\right)^{3/2}\, 2 i \sin(\vec{k}_{\perp}\!\cdot\!\vec{x}_{\perp})\right]\biggl|^{2}\nonumber\\
     &\simeq&\frac{\lambda^{2}\beta f^{2}}{\pi^{3} \hbar_{d} a^{2}}\left[\left(\ln\!\left[\frac{\Delta E \sin\theta}{2\hbar_{d} a}\right]+\gamma\right)^{2}\, \cos^{2}(\vec{k}_{\perp}\!\cdot\!\vec{x}_{\perp})\right]~.
\end{eqnarray}
In obtaining the above result, we have used the large-acceleration expansions $e^{\pm\frac{\pi \Delta E}{2 \hbar_{d} a}}/{\sinh^{1/2}\!\left(\frac{\pi \Delta E}{\hbar_{d} a}\right)}\simeq\sqrt{{\hbar_{d} a}/{\pi \Delta E}}\pm \frac{1}{2}\sqrt{{\pi \Delta E}/{\hbar_{d} a}}$ and $\sin\!\left(\frac{\Delta E}{\hbar_{d}a}\left[\ln\!\left(\frac{\Delta E\,\sin\theta}{2\hbar_{d}a}\right)+\gamma\right]\right)\simeq\frac{\Delta E}{\hbar_{d}a}\left[\ln\!\left(\frac{\Delta E\,\sin\theta}{2\hbar_{d}a}\right)+\gamma\right] $, which are valid for $\Delta E/(\hbar_{d}a)\ll1$.  Furthermore, in the third step, we have neglected the subleading term inside the modulus in comparison with the leading term in the limit $\Delta E/(\hbar a)\ll1$, which leads directly to the final simplified expression.

\subsubsection{A static atom}

For the large acceleration, we have $\omega/a\ll1$ and we can expand $\varphi_{ex}\thickapprox\,-\frac{\omega}{a}\ln{\frac{k_{\perp}}{2\Omega}}$. With this approximation, we can rewrite the effective field contribution to excitation probability (\ref{eq: P-NM-a-e-3s}) as
\begin{eqnarray}
      P_{\alpha}(\Omega)&\simeq\frac{\lambda^2\alpha_{k}^{2}\,\hbar_{f}}{4\pi^2a\Omega^2\,\hbar_{d}\sinh{(\frac{\pi\omega}{a})}}
\big|-e^{-\frac{\pi\omega }{2a}+i\,\vec{k}_{\perp}\cdot\vec{x}_{\perp}}\cos{(\Omega z_0-\frac{\omega}{a}\ln{(\frac{k_{\perp}}{2\Omega})})}+e^{\frac{\pi\omega }{2a}-i\,\vec{k}_{\perp}\cdot\vec{x}_{\perp}}\cos{(\Omega z_0+\frac{\omega}{a}\ln{(\frac{k_{\perp}}{2\Omega})})}\big|^{2}\,.
\end{eqnarray}
The classical-field regime is obtained by taking the limit $\hbar_f\to0$ while keeping the product $\alpha_k^{2}\hbar_f$ fixed, so that the coherent-state amplitude $\alpha_k$ is related to the classical mode amplitude $f(k)$ through the relation $\alpha_k^{2}\hbar=\beta\,\omega f^{2}$. We impose the resonance condition $\Omega=\omega$ and express the detector frequency as $\Omega=\Delta E/\hbar_d$. The large-acceleration regime is then characterized by $\Delta E/(\hbar_d a)\ll1$. In this limit, the effective-field contribution $P_{\alpha}$ can be simplified as follows:
\begin{eqnarray}
      P_{\alpha}(\Omega)
&\simeq&\frac{\lambda^2\beta\,f^{2}}{4\pi^2a\Delta E\,}
\big|-e^{i\vec{k}_{\perp}.\vec{x}_{\perp}}\frac{e^{-\frac{\pi  \Delta E\, }{2\hbar_{d} a}}}{\sinh^{1/2}{\frac{\pi\Delta E\,}{\hbar_{d}a}}}\cos{(\frac{\Delta E\, z_0}{\hbar_{d}}-\frac{\Delta E\,}{\hbar_{d}a}\ln{(\frac{\sin{\theta}}{2})})}+e^{-i\vec{k}_{\perp}.\vec{x}_{\perp}}\frac{e^{\frac{\pi  \Delta E\, }{2\hbar_{d} a}}}{\sinh^{1/2}{\frac{\pi\Delta E\,}{\hbar_{d}a}}}\cos{(\frac{\Delta E\, z_0}{\hbar_{d}}+\frac{\Delta E\,}{\hbar_{d}a}\ln{(\frac{\sin{\theta}}{2})})}\big|^{2}\,\nonumber\\
&\simeq&\frac{\lambda^2\beta\,f^{2}}{4\pi^2a\Delta E\,}
\big|-e^{i\vec{k}_{\perp}.\vec{x}_{\perp}}\frac{e^{-\frac{\pi  \Delta E\, }{2\hbar_{d} a}}}{\sinh^{1/2}{\frac{\pi\Delta E\,}{\hbar_{d}a}}}\cos{(\frac{\Delta E\,}{\hbar_{d}a}\ln{(\frac{\sin{\theta}}{2})})} \,+\,e^{-i\vec{k}_{\perp}.\vec{x}_{\perp}}\frac{e^{\frac{\pi  \Delta E\, }{2\hbar_{d} a}}}{\sinh^{1/2}{\frac{\pi\Delta E\,}{\hbar_{d}a}}}\cos{(\frac{\Delta E\,}{\hbar_{d}a}\ln{(\frac{\sin{\theta}}{2})})}\big|^{2}\,\nonumber\\
&\simeq&\frac{\lambda^2 \beta f^2}{4\pi^2 a \Delta E}\left|\sqrt{\frac{\hbar_{d} a}{\pi \Delta E}}\, 2 i \sin(\vec{k}_{\perp}\!\cdot\!\vec{x}_{\perp})+ \frac{1}{2}\sqrt{\frac{\pi \Delta E}{\hbar_{d} a}}\cos(\vec{k}_{\perp}\!\cdot\!\vec{x}_{\perp})\right|^2\,\nonumber\\
&\simeq&\frac{\lambda^2 \beta f^2\,\hbar_{d}}{\pi^3  \Delta E^{2}}\, \sin^{2}(\vec{k}_{\perp}\!\cdot\!\vec{x}_{\perp})\,~.
\end{eqnarray}
In obtaining the above result, in the first step, we have used $k_{\perp}=\omega\sin\theta$. We have also employed the large-acceleration expansions $e^{\pm\frac{\pi \Delta E}{2 \hbar_{d} a}}/{\sinh^{1/2}\!\left(\frac{\pi \Delta E}{\hbar_{d} a}\right)}\simeq\sqrt{{\hbar_{d} a}/{\pi \Delta E}}\pm \frac{1}{2}\sqrt{{\pi \Delta E}/{\hbar_{d} a}}$ and $\cos\!\left(\frac{\Delta E}{\hbar_{d} a}\ln\!\left(\frac{\sin\theta}{2}\right)\right)\simeq \,1$, which are valid for $\Delta E/(\hbar_{d}a)\ll1$. Furthermore, in the second step, we have set $z_{0}=0$, and in the third step, we have neglected the subleading term inside the modulus in comparison with the leading term in the limit $\Delta E/(\hbar a)\ll1$, which directly yields the final simplified expression.


\end{appendix}
\end{widetext}

\bibliographystyle{apsrev4-2}

\bibliography{bibtexfile}

\begin{thebibliography}{27}%
\makeatletter
\providecommand \@ifxundefined [1]{%
 \@ifx{#1\undefined}
}%
\providecommand \@ifnum [1]{%
 \ifnum #1\expandafter \@firstoftwo
 \else \expandafter \@secondoftwo
 \fi
}%
\providecommand \@ifx [1]{%
 \ifx #1\expandafter \@firstoftwo
 \else \expandafter \@secondoftwo
 \fi
}%
\providecommand \natexlab [1]{#1}%
\providecommand \enquote  [1]{``#1''}%
\providecommand \bibnamefont  [1]{#1}%
\providecommand \bibfnamefont [1]{#1}%
\providecommand \citenamefont [1]{#1}%
\providecommand \href@noop [0]{\@secondoftwo}%
\providecommand \href [0]{\begingroup \@sanitize@url \@href}%
\providecommand \@href[1]{\@@startlink{#1}\@@href}%
\providecommand \@@href[1]{\endgroup#1\@@endlink}%
\providecommand \@sanitize@url [0]{\catcode `\\12\catcode `\$12\catcode `\&12\catcode `\#12\catcode `\^12\catcode `\_12\catcode `\%12\relax}%
\providecommand \@@startlink[1]{}%
\providecommand \@@endlink[0]{}%
\providecommand \url  [0]{\begingroup\@sanitize@url \@url }%
\providecommand \@url [1]{\endgroup\@href {#1}{\urlprefix }}%
\providecommand \urlprefix  [0]{URL }%
\providecommand \Eprint [0]{\href }%
\providecommand \doibase [0]{https://doi.org/}%
\providecommand \selectlanguage [0]{\@gobble}%
\providecommand \bibinfo  [0]{\@secondoftwo}%
\providecommand \bibfield  [0]{\@secondoftwo}%
\providecommand \translation [1]{[#1]}%
\providecommand \BibitemOpen [0]{}%
\providecommand \bibitemStop [0]{}%
\providecommand \bibitemNoStop [0]{.\EOS\space}%
\providecommand \EOS [0]{\spacefactor3000\relax}%
\providecommand \BibitemShut  [1]{\csname bibitem#1\endcsname}%
\let\auto@bib@innerbib\@empty
\bibitem [{\citenamefont {Unruh}(1976)}]{Unruh:1976db}%
  \BibitemOpen
  \bibfield  {author} {\bibinfo {author} {\bibfnamefont {W.~G.}\ \bibnamefont {Unruh}},\ }\href {https://doi.org/10.1103/PhysRevD.14.870} {\bibfield  {journal} {\bibinfo  {journal} {Phys. Rev. D}\ }\textbf {\bibinfo {volume} {14}},\ \bibinfo {pages} {870} (\bibinfo {year} {1976})}\BibitemShut {NoStop}%
\bibitem [{\citenamefont {HAWKING}(1974)}]{HAWKING:1974us}%
  \BibitemOpen
  \bibfield  {author} {\bibinfo {author} {\bibfnamefont {S.~W.}\ \bibnamefont {HAWKING}},\ }\href {https://doi.org/10.1038/248030a0} {\bibfield  {journal} {\bibinfo  {journal} {Nature}\ }\textbf {\bibinfo {volume} {248}},\ \bibinfo {pages} {30} (\bibinfo {year} {1974})}\BibitemShut {NoStop}%
\bibitem [{\citenamefont {Hawking}(1975)}]{Hawking1975}%
  \BibitemOpen
  \bibfield  {author} {\bibinfo {author} {\bibfnamefont {S.~W.}\ \bibnamefont {Hawking}},\ }\href@noop {} {\bibfield  {journal} {\bibinfo  {journal} {Communications in Mathematical Physics}\ }\textbf {\bibinfo {volume} {43}},\ \bibinfo {pages} {199} (\bibinfo {year} {1975})}\BibitemShut {NoStop}%
\bibitem [{\citenamefont {Singleton}\ and\ \citenamefont {Wilburn}(2011)}]{Singleton:2011vh}%
  \BibitemOpen
  \bibfield  {author} {\bibinfo {author} {\bibfnamefont {D.}~\bibnamefont {Singleton}}\ and\ \bibinfo {author} {\bibfnamefont {S.}~\bibnamefont {Wilburn}},\ }\href {https://doi.org/10.1103/PhysRevLett.107.081102} {\bibfield  {journal} {\bibinfo  {journal} {Phys. Rev. Lett.}\ }\textbf {\bibinfo {volume} {107}},\ \bibinfo {pages} {081102} (\bibinfo {year} {2011})},\ \Eprint {https://arxiv.org/abs/arXiv:1102.5564} {arXiv:arXiv:1102.5564 [gr-qc]} \BibitemShut {NoStop}%
\bibitem [{\citenamefont {Zych}\ and\ \citenamefont {Brukner}(2018)}]{Zych:2015fka}%
  \BibitemOpen
  \bibfield  {author} {\bibinfo {author} {\bibfnamefont {M.}~\bibnamefont {Zych}}\ and\ \bibinfo {author} {\bibfnamefont {{\v{C}}.}~\bibnamefont {Brukner}},\ }\href {https://doi.org/10.1038/s41567-018-0197-6} {\bibfield  {journal} {\bibinfo  {journal} {Nature Phys.}\ }\textbf {\bibinfo {volume} {14}},\ \bibinfo {pages} {1027} (\bibinfo {year} {2018})},\ \Eprint {https://arxiv.org/abs/1502.00971} {arXiv:1502.00971 [gr-qc]} \BibitemShut {NoStop}%
\bibitem [{\citenamefont {Paunkovic}\ and\ \citenamefont {Vojinovic}(2022)}]{Paunkovic:2022flx}%
  \BibitemOpen
  \bibfield  {author} {\bibinfo {author} {\bibfnamefont {N.}~\bibnamefont {Paunkovic}}\ and\ \bibinfo {author} {\bibfnamefont {M.}~\bibnamefont {Vojinovic}},\ }\href {https://doi.org/10.3390/universe8110598} {\bibfield  {journal} {\bibinfo  {journal} {Universe}\ }\textbf {\bibinfo {volume} {8}},\ \bibinfo {pages} {598} (\bibinfo {year} {2022})},\ \Eprint {https://arxiv.org/abs/2210.00133} {arXiv:2210.00133 [gr-qc]} \BibitemShut {NoStop}%
\bibitem [{\citenamefont {Svidzinsky}\ \emph {et~al.}(2018)\citenamefont {Svidzinsky}, \citenamefont {Ben-Benjamin}, \citenamefont {Fulling},\ and\ \citenamefont {Page}}]{PhysRevLett.121.071301}%
  \BibitemOpen
  \bibfield  {author} {\bibinfo {author} {\bibfnamefont {A.~A.}\ \bibnamefont {Svidzinsky}}, \bibinfo {author} {\bibfnamefont {J.~S.}\ \bibnamefont {Ben-Benjamin}}, \bibinfo {author} {\bibfnamefont {S.~A.}\ \bibnamefont {Fulling}},\ and\ \bibinfo {author} {\bibfnamefont {D.~N.}\ \bibnamefont {Page}},\ }\href {https://doi.org/10.1103/PhysRevLett.121.071301} {\bibfield  {journal} {\bibinfo  {journal} {Phys. Rev. Lett.}\ }\textbf {\bibinfo {volume} {121}},\ \bibinfo {pages} {071301} (\bibinfo {year} {2018})}\BibitemShut {NoStop}%
\bibitem [{\citenamefont {Louko}\ and\ \citenamefont {Satz}(2008)}]{Louko:2007mu}%
  \BibitemOpen
  \bibfield  {author} {\bibinfo {author} {\bibfnamefont {J.}~\bibnamefont {Louko}}\ and\ \bibinfo {author} {\bibfnamefont {A.}~\bibnamefont {Satz}},\ }\href {https://doi.org/10.1088/0264-9381/25/5/055012} {\bibfield  {journal} {\bibinfo  {journal} {Class. Quant. Grav.}\ }\textbf {\bibinfo {volume} {25}},\ \bibinfo {pages} {055012} (\bibinfo {year} {2008})},\ \Eprint {https://arxiv.org/abs/0710.5671} {arXiv:0710.5671 [gr-qc]} \BibitemShut {NoStop}%
\bibitem [{\citenamefont {Fulling}\ and\ \citenamefont {Wilson}(2019)}]{Fulling:2018lez}%
  \BibitemOpen
  \bibfield  {author} {\bibinfo {author} {\bibfnamefont {S.~A.}\ \bibnamefont {Fulling}}\ and\ \bibinfo {author} {\bibfnamefont {J.~H.}\ \bibnamefont {Wilson}},\ }\href {https://doi.org/10.1088/1402-4896/aaecaa} {\bibfield  {journal} {\bibinfo  {journal} {Phys. Scripta}\ }\textbf {\bibinfo {volume} {94}},\ \bibinfo {pages} {014004} (\bibinfo {year} {2019})},\ \Eprint {https://arxiv.org/abs/1805.01013} {arXiv:1805.01013 [quant-ph]} \BibitemShut {NoStop}%
\bibitem [{\citenamefont {Das}\ \emph {et~al.}(2023)\citenamefont {Das}, \citenamefont {Fridman},\ and\ \citenamefont {Lambiase}}]{Das:2023cfu}%
  \BibitemOpen
  \bibfield  {author} {\bibinfo {author} {\bibfnamefont {S.}~\bibnamefont {Das}}, \bibinfo {author} {\bibfnamefont {M.}~\bibnamefont {Fridman}},\ and\ \bibinfo {author} {\bibfnamefont {G.}~\bibnamefont {Lambiase}},\ }\href {https://doi.org/10.1038/s42005-023-01306-w} {\bibfield  {journal} {\bibinfo  {journal} {Commun. Phys.}\ }\textbf {\bibinfo {volume} {6}},\ \bibinfo {pages} {198} (\bibinfo {year} {2023})},\ \Eprint {https://arxiv.org/abs/2307.09632} {arXiv:2307.09632 [gr-qc]} \BibitemShut {NoStop}%
\bibitem [{\citenamefont {Barman}\ \emph {et~al.}(2026)\citenamefont {Barman}, \citenamefont {Kumawat},\ and\ \citenamefont {Majhi}}]{Barman:2024dql}%
  \BibitemOpen
  \bibfield  {author} {\bibinfo {author} {\bibfnamefont {S.}~\bibnamefont {Barman}}, \bibinfo {author} {\bibfnamefont {P.~K.}\ \bibnamefont {Kumawat}},\ and\ \bibinfo {author} {\bibfnamefont {B.~R.}\ \bibnamefont {Majhi}},\ }\href {https://doi.org/10.1088/1475-7516/2026/02/055} {\bibfield  {journal} {\bibinfo  {journal} {JCAP}\ }\textbf {\bibinfo {volume} {02}},\ \bibinfo {pages} {055}},\ \Eprint {https://arxiv.org/abs/2408.12378} {arXiv:2408.12378 [gr-qc]} \BibitemShut {NoStop}%
\bibitem [{\citenamefont {Kumawat}\ \emph {et~al.}(2025)\citenamefont {Kumawat}, \citenamefont {Barman},\ and\ \citenamefont {Majhi}}]{Kumawat:2024kul}%
  \BibitemOpen
  \bibfield  {author} {\bibinfo {author} {\bibfnamefont {P.~K.}\ \bibnamefont {Kumawat}}, \bibinfo {author} {\bibfnamefont {S.}~\bibnamefont {Barman}},\ and\ \bibinfo {author} {\bibfnamefont {B.~R.}\ \bibnamefont {Majhi}},\ }\href {https://doi.org/10.1088/1475-7516/2025/02/046} {\bibfield  {journal} {\bibinfo  {journal} {JCAP}\ }\textbf {\bibinfo {volume} {02}},\ \bibinfo {pages} {046}},\ \Eprint {https://arxiv.org/abs/2410.21051} {arXiv:2410.21051 [gr-qc]} \BibitemShut {NoStop}%
\bibitem [{\citenamefont {Padmanabhan}(2010)}]{book:PadmanabhanGrav}%
  \BibitemOpen
  \bibfield  {author} {\bibinfo {author} {\bibfnamefont {T.}~\bibnamefont {Padmanabhan}},\ }\href@noop {} {\emph {\bibinfo {title} {Gravitation: Foundations and Frontiers}}},\ \bibinfo {edition} {1st}\ ed.\ (\bibinfo  {publisher} {Cambridge University Press},\ \bibinfo {year} {2010})\BibitemShut {NoStop}%
\bibitem [{\citenamefont {Singh}\ \emph {et~al.}(2013)\citenamefont {Singh}, \citenamefont {Ganguly},\ and\ \citenamefont {Padmanabhan}}]{Singh:2013pxf}%
  \BibitemOpen
  \bibfield  {author} {\bibinfo {author} {\bibfnamefont {S.}~\bibnamefont {Singh}}, \bibinfo {author} {\bibfnamefont {C.}~\bibnamefont {Ganguly}},\ and\ \bibinfo {author} {\bibfnamefont {T.}~\bibnamefont {Padmanabhan}},\ }\href {https://doi.org/10.1103/PhysRevD.87.104004} {\bibfield  {journal} {\bibinfo  {journal} {Phys. Rev.}\ }\textbf {\bibinfo {volume} {D87}},\ \bibinfo {pages} {104004} (\bibinfo {year} {2013})},\ \Eprint {https://arxiv.org/abs/arXiv:1302.7177} {arXiv:arXiv:1302.7177 [gr-qc]} \BibitemShut {NoStop}%
\bibitem [{\citenamefont {Padmanabhan}\ and\ \citenamefont {Singh}(1988)}]{PhysRevD.38.2457}%
  \BibitemOpen
  \bibfield  {author} {\bibinfo {author} {\bibfnamefont {T.}~\bibnamefont {Padmanabhan}}\ and\ \bibinfo {author} {\bibfnamefont {T.~P.}\ \bibnamefont {Singh}},\ }\href {https://doi.org/10.1103/PhysRevD.38.2457} {\bibfield  {journal} {\bibinfo  {journal} {Phys. Rev. D}\ }\textbf {\bibinfo {volume} {38}},\ \bibinfo {pages} {2457} (\bibinfo {year} {1988})}\BibitemShut {NoStop}%
\bibitem [{\citenamefont {Lochan}\ and\ \citenamefont {Padmanabhan}(2015)}]{Lochan:2014xja}%
  \BibitemOpen
  \bibfield  {author} {\bibinfo {author} {\bibfnamefont {K.}~\bibnamefont {Lochan}}\ and\ \bibinfo {author} {\bibfnamefont {T.}~\bibnamefont {Padmanabhan}},\ }\href {https://doi.org/10.1103/PhysRevD.91.044002} {\bibfield  {journal} {\bibinfo  {journal} {Phys. Rev. D}\ }\textbf {\bibinfo {volume} {91}},\ \bibinfo {pages} {044002} (\bibinfo {year} {2015})},\ \Eprint {https://arxiv.org/abs/arXiv:1411.7019} {arXiv:arXiv:1411.7019 [gr-qc]} \BibitemShut {NoStop}%
\bibitem [{\citenamefont {Barman}\ \emph {et~al.}(2022)\citenamefont {Barman}, \citenamefont {Barman},\ and\ \citenamefont {Majhi}}]{Barman:2022xht}%
  \BibitemOpen
  \bibfield  {author} {\bibinfo {author} {\bibfnamefont {D.}~\bibnamefont {Barman}}, \bibinfo {author} {\bibfnamefont {S.}~\bibnamefont {Barman}},\ and\ \bibinfo {author} {\bibfnamefont {B.~R.}\ \bibnamefont {Majhi}},\ }\href {https://doi.org/10.1103/PhysRevD.106.045005} {\bibfield  {journal} {\bibinfo  {journal} {Phys. Rev. D}\ }\textbf {\bibinfo {volume} {106}},\ \bibinfo {pages} {045005} (\bibinfo {year} {2022})},\ \Eprint {https://arxiv.org/abs/2205.08505} {arXiv:2205.08505 [gr-qc]} \BibitemShut {NoStop}%
\bibitem [{\citenamefont {Bruschi}\ \emph {et~al.}(2010)\citenamefont {Bruschi}, \citenamefont {Louko}, \citenamefont {Martin-Martinez}, \citenamefont {Dragan},\ and\ \citenamefont {Fuentes}}]{Bruschi:2010mc}%
  \BibitemOpen
  \bibfield  {author} {\bibinfo {author} {\bibfnamefont {D.~E.}\ \bibnamefont {Bruschi}}, \bibinfo {author} {\bibfnamefont {J.}~\bibnamefont {Louko}}, \bibinfo {author} {\bibfnamefont {E.}~\bibnamefont {Martin-Martinez}}, \bibinfo {author} {\bibfnamefont {A.}~\bibnamefont {Dragan}},\ and\ \bibinfo {author} {\bibfnamefont {I.}~\bibnamefont {Fuentes}},\ }\href {https://doi.org/10.1103/PhysRevA.82.042332} {\bibfield  {journal} {\bibinfo  {journal} {Phys. Rev. A}\ }\textbf {\bibinfo {volume} {82}},\ \bibinfo {pages} {042332} (\bibinfo {year} {2010})},\ \Eprint {https://arxiv.org/abs/1007.4670} {arXiv:1007.4670 [quant-ph]} \BibitemShut {NoStop}%
\bibitem [{\citenamefont {Scully}\ \emph {et~al.}(2018)\citenamefont {Scully}, \citenamefont {Fulling}, \citenamefont {Lee}, \citenamefont {Page}, \citenamefont {Schleich},\ and\ \citenamefont {Svidzinsky}}]{Scully:2017utk}%
  \BibitemOpen
  \bibfield  {author} {\bibinfo {author} {\bibfnamefont {M.~O.}\ \bibnamefont {Scully}}, \bibinfo {author} {\bibfnamefont {S.}~\bibnamefont {Fulling}}, \bibinfo {author} {\bibfnamefont {D.}~\bibnamefont {Lee}}, \bibinfo {author} {\bibfnamefont {D.~N.}\ \bibnamefont {Page}}, \bibinfo {author} {\bibfnamefont {W.}~\bibnamefont {Schleich}},\ and\ \bibinfo {author} {\bibfnamefont {A.}~\bibnamefont {Svidzinsky}},\ }\href {https://doi.org/10.1073/pnas.1807703115} {\bibfield  {journal} {\bibinfo  {journal} {Proc. Nat. Acad. Sci.}\ }\textbf {\bibinfo {volume} {115}},\ \bibinfo {pages} {8131} (\bibinfo {year} {2018})},\ \Eprint {https://arxiv.org/abs/arXiv:1709.00481} {arXiv:arXiv:1709.00481 [quant-ph]} \BibitemShut {NoStop}%
\bibitem [{\citenamefont {Chakraborty}\ and\ \citenamefont {Majhi}(2019)}]{Chakraborty:2019ltu}%
  \BibitemOpen
  \bibfield  {author} {\bibinfo {author} {\bibfnamefont {K.}~\bibnamefont {Chakraborty}}\ and\ \bibinfo {author} {\bibfnamefont {B.~R.}\ \bibnamefont {Majhi}},\ }\href {https://doi.org/10.1103/PhysRevD.100.045004} {\bibfield  {journal} {\bibinfo  {journal} {Phys. Rev. D}\ }\textbf {\bibinfo {volume} {100}},\ \bibinfo {pages} {045004} (\bibinfo {year} {2019})},\ \Eprint {https://arxiv.org/abs/arXiv:1905.10554} {arXiv:arXiv:1905.10554 [gr-qc]} \BibitemShut {NoStop}%
\bibitem [{\citenamefont {Olson}\ and\ \citenamefont {Ralph}(2011)}]{Olson:2010jy}%
  \BibitemOpen
  \bibfield  {author} {\bibinfo {author} {\bibfnamefont {S.~J.}\ \bibnamefont {Olson}}\ and\ \bibinfo {author} {\bibfnamefont {T.~C.}\ \bibnamefont {Ralph}},\ }\href {https://doi.org/10.1103/PhysRevLett.106.110404} {\bibfield  {journal} {\bibinfo  {journal} {Phys. Rev. Lett.}\ }\textbf {\bibinfo {volume} {106}},\ \bibinfo {pages} {110404} (\bibinfo {year} {2011})},\ \Eprint {https://arxiv.org/abs/1003.0720} {arXiv:1003.0720 [quant-ph]} \BibitemShut {NoStop}%
\bibitem [{\citenamefont {Higuchi}\ \emph {et~al.}(2017)\citenamefont {Higuchi}, \citenamefont {Iso}, \citenamefont {Ueda},\ and\ \citenamefont {Yamamoto}}]{Higuchi:2017gcd}%
  \BibitemOpen
  \bibfield  {author} {\bibinfo {author} {\bibfnamefont {A.}~\bibnamefont {Higuchi}}, \bibinfo {author} {\bibfnamefont {S.}~\bibnamefont {Iso}}, \bibinfo {author} {\bibfnamefont {K.}~\bibnamefont {Ueda}},\ and\ \bibinfo {author} {\bibfnamefont {K.}~\bibnamefont {Yamamoto}},\ }\href {https://doi.org/10.1103/PhysRevD.96.083531} {\bibfield  {journal} {\bibinfo  {journal} {Phys. Rev. D}\ }\textbf {\bibinfo {volume} {96}},\ \bibinfo {pages} {083531} (\bibinfo {year} {2017})},\ \Eprint {https://arxiv.org/abs/arXiv:1709.05757} {arXiv:arXiv:1709.05757 [hep-th]} \BibitemShut {NoStop}%
\bibitem [{\citenamefont {Gradshteyn}\ and\ \citenamefont {Ryzhik}(2014)}]{Gradshteyn}%
  \BibitemOpen
  \bibfield  {author} {\bibinfo {author} {\bibfnamefont {I.~S.}\ \bibnamefont {Gradshteyn}}\ and\ \bibinfo {author} {\bibfnamefont {I.~M.}\ \bibnamefont {Ryzhik}},\ }\href@noop {} {\emph {\bibinfo {title} {Table of Integrals, Series, and Products}}}\ (\bibinfo  {publisher} {Academic Press},\ \bibinfo {year} {2014})\BibitemShut {NoStop}%
\bibitem [{\citenamefont {Boyer}(1980)}]{PhysRevD.21.2137}%
  \BibitemOpen
  \bibfield  {author} {\bibinfo {author} {\bibfnamefont {T.~H.}\ \bibnamefont {Boyer}},\ }\href {https://doi.org/10.1103/PhysRevD.21.2137} {\bibfield  {journal} {\bibinfo  {journal} {Phys. Rev. D}\ }\textbf {\bibinfo {volume} {21}},\ \bibinfo {pages} {2137} (\bibinfo {year} {1980})}\BibitemShut {NoStop}%
\bibitem [{\citenamefont {Crispino}\ \emph {et~al.}(2008)\citenamefont {Crispino}, \citenamefont {Higuchi},\ and\ \citenamefont {Matsas}}]{Crispino:2007eb}%
  \BibitemOpen
  \bibfield  {author} {\bibinfo {author} {\bibfnamefont {L.~C.~B.}\ \bibnamefont {Crispino}}, \bibinfo {author} {\bibfnamefont {A.}~\bibnamefont {Higuchi}},\ and\ \bibinfo {author} {\bibfnamefont {G.~E.~A.}\ \bibnamefont {Matsas}},\ }\href {https://doi.org/10.1103/RevModPhys.80.787} {\bibfield  {journal} {\bibinfo  {journal} {Rev. Mod. Phys.}\ }\textbf {\bibinfo {volume} {80}},\ \bibinfo {pages} {787} (\bibinfo {year} {2008})},\ \Eprint {https://arxiv.org/abs/0710.5373} {arXiv:0710.5373 [gr-qc]} \BibitemShut {NoStop}%
\bibitem [{\citenamefont {Olver}(1997)}]{olver1997asymptotics}%
  \BibitemOpen
  \bibfield  {author} {\bibinfo {author} {\bibfnamefont {F.}~\bibnamefont {Olver}},\ }\href@noop {} {\emph {\bibinfo {title} {Asymptotics and special functions}}}\ (\bibinfo  {publisher} {AK Peters/CRC Press},\ \bibinfo {year} {1997})\BibitemShut {NoStop}%
\bibitem [{\citenamefont {Dunster}(1990)}]{dunster1990bessel}%
  \BibitemOpen
  \bibfield  {author} {\bibinfo {author} {\bibfnamefont {T.~M.}\ \bibnamefont {Dunster}},\ }\href@noop {} {\bibfield  {journal} {\bibinfo  {journal} {SIAM Journal on Mathematical Analysis}\ }\textbf {\bibinfo {volume} {21}},\ \bibinfo {pages} {995} (\bibinfo {year} {1990})}\BibitemShut {NoStop}%
\end{thebibliography}%

\end{document}